\documentclass
[twocolumn,showpacs,preprintnumbers,amsmath,amssymb,aps,pre,superscriptaddress,floatfix,reprint]
{revtex4}
\usepackage{graphicx}
\usepackage{dcolumn}
\usepackage{bm}
\usepackage{here}
\usepackage{color} 
\usepackage{braket}
\usepackage{amsmath}  
\usepackage{amssymb}  
\usepackage{fancybox}

\begin{document}

\title{Localization and delocalization properties in quasi-periodically 
driven one-dimensional disordered system}
\author{Hiroaki S. Yamada}
\affiliation{Yamada Physics Research Laboratory,
Aoyama 5-7-14-205, Niigata 950-2002, Japan}
\author{Kensuke S. Ikeda}
\affiliation{College of Science and Engineering, Ritsumeikan University, 
Noji-higashi 1-1-1, Kusatsu 525-8577, Japan}

\date{\today}

\newcommand{\vc}[1]{\mbox{\boldmath $#1$}}
\newcommand{\fracd}[2]{\frac{\displaystyle #1}{\displaystyle #2}}
\newcommand{\red}[1]{\textcolor{red}{#1}}
\newcommand{\blue}[1]{\textcolor{blue}{#1}}
\newcommand{\green}[1]{\textcolor{green}{#1}}

\begin{abstract}
Localization and delocalization of quantum diffusion 
in time-continuous one-dimensional Anderson model 
perturbed by the quasi-periodic harmonic oscillations of $M$ colors 
is investigated systematically, which has been partly reported by the 
preliminary letter [PRE {\bf 103}, L040202(2021)].
We investigate in detail the localization-delocalization characteristics of the model 
with respect to three parameters: the disorder strength $W$, the perturbation strength $\epsilon$
and the number of the colors $M$ which plays the similar role of spatial dimension. 
In particular, attentions are focused on the presence of localization-delocalization 
transition (LDT) and its critical properties. 
For $M\geq 3$ the LDT exists and a normal diffusion is recovered above a 
critical strength  $\epsilon$, and the characteristics of diffusion dynamics mimic the diffusion process 
predicted for the stochastically perturbed Anderson model even though $M$ is not large. These results are 
compared with the results of time-discrete quantum maps, ie., Anderson map and 
the standard map. Further, the features of delocalized dynamics is discussed in comparison 
with a limit model which has no static disordered part. 
\end{abstract}



\maketitle


\newcommand{\del}{\partial}

\def\ni{\noindent}
\def\nn{\nonumber}
\def\bH{\begin{Huge}}
\def\eH{\end{Huge}}
\def\bL{\begin{Large}}
\def\eL{\end{Large}}
\def\bl{\begin{large}}
\def\el{\end{large}}
\def\beq{\begin{eqnarray}}
\def\eeq{\end{eqnarray}}
\def\beqnn{\begin{eqnarray*}}
\def\eeqnn{\end{eqnarray*}}

\def\bit{\begin{itemize}}
\def\eit{\end{itemize}}
\def\bsc{\begin{screen}}
\def\esc{\end{screen}}

\def\eps{\epsilon}
\def\th{\theta}
\def\del{\delta}
\def\omg{\omega}

\def\e{{\rm e}}
\def\exp{{\rm exp}}
\def\arg{{\rm arg}}
\def\Im{{\rm Im}}
\def\Re{{\rm Re}}

\def\sup{\supset}
\def\sub{\subset}
\def\a{\cap}
\def\u{\cup}
\def\bks{\backslash}

\def\ovl{\overline}
\def\unl{\underline}

\def\rar{\rightarrow}
\def\Rar{\Rightarrow}
\def\lar{\leftarrow}
\def\Lar{\Leftarrow}
\def\bar{\leftrightarrow}
\def\Bar{\Leftrightarrow}

\def\pr{\partial}

\def\>{\rangle} 
\def\<{\langle} 
\def\RR {\rangle\!\rangle} 
\def\LL {\langle\!\langle} 
\def\const{{\rm const.}}

\def\e{{\rm e}}

\def\Bstar{\bL $\star$ \eL}

\def\etath{\eta_{th}}
\def\irrev{{\mathcal R}}
\def\e{{\rm e}}
\def\noise{n}
\def\hatp{\hat{p}}
\def\hatq{\hat{q}}
\def\hatU{\hat{U}}

\def\hatA{\hat{A}}
\def\hatB{\hat{B}}
\def\hatC{\hat{C}}
\def\hatJ{\hat{J}}
\def\hatI{\hat{I}}
\def\hatP{\hat{P}}
\def\hatQ{\hat{Q}}
\def\hatU{\hat{U}}
\def\hatW{\hat{W}}
\def\hatX{\hat{X}}
\def\hatY{\hat{Y}}
\def\hatV{\hat{V}}
\def\hatt{\hat{t}}
\def\hatw{\hat{w}}

\def\hatp{\hat{p}}
\def\hatq{\hat{q}}
\def\hatU{\hat{U}}
\def\hatn{\hat{n}}

\def\hatphi{\hat{\phi}}
\def\hattheta{\hat{\theta}}

\def\iset{\mathcal{I}}
\def\fset{\mathcal{F}}
\def\pr{\partial}
\def\traj{\ell}
\def\eps{\epsilon}
\def\U{\hat{U}}

\def\U{U_{\rm cls}}
\def\P{P_{{\rm cls},\eta}}
\def\traj{\ell}
\def\cc{\cdot}

\def\DZ{D^{(0)}}
\def\Dcls{D_{\rm cls}}

\def\deff{d_f}
\def\alphac{\alpha}
\def\alphac{\alpha_{\rm ins}}

\newcommand{\relmiddle}[1]{\mathrel{}\middle#1\mathrel{}}

\section{Introduction}
It has been theoretically and experimentally shown that the  three-dimensional random system 
undergoes an Anderson transition (AT) from insulator to metallic conductor 
 due to decrease in the potential disorder \cite{anderson58,ishii73,lifshiz88,abrahams10}. 
Furthermore, in recent numerical experiments, 
the properties of AT in 4-dimensional and 5-dimensional random systems 
have been also studied \cite{markos06,garcia07,slevin14,tarquini17}. 
In the system with the  AT, a localization-delocalization transition (LDT) can exist, 
and its existence can be directly observed by the wavepacket dynamics of 
initially localized wave packet, where the delocalization is observed
as an appearance of normal diffusion.

In higher-dimensional Anderson model the appearance of delocalized
states is quite natural, and it is expected that the self-consistent 
mean-field theory works well in such systems \cite{vollhard80,wolfle10}. 
However, even in higher-dimensional Anderson models the deviation of 
the critical value and the critical exponent predicted  by the SCT was 
recently reported by using the properties of the energy spectrum \cite{tarquini17}. 

The relationship between the dimension of the Anderson model and the characteristics 
of the LDT is an interesting problem from a different point of view.
Increase of the system's dimension $d$ may be performed in a quite different way:
an alternative way to increase $d$ is to make the 
system interact with many dynamical degrees of freedom.
Indeed, even in the one-dimensional  (1D) Anderson model
exhibiting a strong exponential localization,  
the localization is released and normal diffusion is induced by the application of 
arbitrarily small stochastic perturbation, which can be considered as
a superposition of an infinite number of incommensurate harmonic degrees of freedoms 
\cite{haken72,palenberg00,moix13,knap17}.  
This can be considered as a limiting example of delocalization realized in systems with
infinite degrees of freedom.

Then it is a quite natural question to inquire how the number of the degrees of 
harmonic modes $M$ controlls the localization and delocalization in disordered
systems. (The harmonic modes may be replaced the active phonon modes.) 
Indeed, in the case of chaotic quantum maps such as the standard map (SM), the harmonic
perturbation destroys the dynamical localization and restores the chaotic diffusion
\cite{casati89,lopez12,lopez13,yamada15,yamada18,yamada20}, which is supported by the Maryland transformation
asserting the equivalence between the SM and a $M+1$-dimensional lattice with
a quasi-periodic disorder.

Quantum maps is a very powerful model which can easily be treated by numerical method
because its time is discretized, but it is not a natural system.
Instead, as a time-continuous model, we proposed a time-continuous 
1D  Anderson model interacting with $M$ 
incommensurate harmonic modes \cite{yamada98,yamada99}. 
For $M=1$, the maintenance of localization can be
shown by the Floquet theory \cite{holhaus95,Martinez06}. But for $M \geq 2$, diffusion-like 
behaviors are observed numerically at least on a finite time scale if the perturbation 
strength is strong enough.
In this system, the $M-$modes can be treated as a quantum dynamical degrees of freedom, 
and so the whole system can be regarded as an autonomous quantum dynamical system with
$(M+1)-$degrees of freedom. 
There has been some studies showing strong localized property of the dynamics for 
the same type of harmonically perturbed models. It is inferred from analytical calculation 
and rigorous proofs that the localization persists against the dynamical perturbation 
consisting of finite number of the modes \cite{hatami16,wang04}.
In particular, the persistence of the localization for $1 \leq M < \infty$ 
is mathematically claimed in the regime of weak 
enough dynamical perturbations and strong disorder potential \cite{wang04}.
On the other hand, as mentioned above, a stochastic perturbation which 
corresponds to $M\to\infty$ can restore a complete diffusion.
The presence of the LDT in harmonically perturbed 1D Anderson model has not been
yet clarified.

In our preliminary report it was shown that if there exist three or more harmonics ($M\geq 3$), 
the LDT occurs with the increase of the perturbation strength and 
the Anderson localized states can be delocalized \cite{yamada21}.
This work is a full report of the localization-delocalization
characteristics of the 1D Anderson model perturbed by polychromatic perturbations, 
which is numerically observed by changing the three parameters: the disorder strength $W$, 
perturbation strength $\epsilon$ and the number of the modes $M$ of the oscillations.
We are particularly interested in making clear how the number $M$ controls 
the characteristics of LDT.  
Additionally as a limiting situation of our model mentioned above, we can consider a model 
system without the static 
random potential. Such a version leads to a quantum state that models the ultimate limit 
of delocalization exhibited by our model, which will be discussed in detail.

Since the direct numerical wavepacket propagation of the original continuous-time model
is too time consuming, we proposed a discrete-time quantum map version of the 
original time-continuous model, which we called the Anderson map (AM), and 
investigated its nature in comparison with the SM and many-dimensional
Anderson model \cite{yamada15,yamada18,yamada20}. Comparison of the
original time-continuous model with the AM is also a purpose of this article.

Recently realization of ergodic state in isolated quantum systems with many degrees of
freedom has been extensively studied \cite{gutzwiller91,borgonovi97,neill16,notarnicola18,piga19}.
As mentioned above, our system is a closed quantum dynamical system with $M+1$
degrees of freedom, and the LDT may be looked upon as a transition to an ergodic
state even though $M$ is small. The transition to a delocalized behavior is a 
``self-organization'' of a irreversible relaxation process in quantum systems with a small-number 
of degrees of freedom stressed in Ref.\cite{ikeda93}.
With this regard the minimal number of $M$ above which the LDT takes place is 
a quite interesting problem.

The plan of the present work is as follows.
In the next section, the models used in the present paper are introduced.
In Sec.\ref{sec:localization}, the characteristics of the localization phase 
which is dominant when the number $M$ is small ie., $M=0,1,2$ are explored. 
A hypothesis due to the intrinsic nature of time-continuous model, 
 which was not taken into account in our preliminary report \cite{yamada21},
is discussed. It is used as a base of the following analysis.
Next, in Sec.\ref{sec:LDT}, the presence of LDT for the case of $M \geq 3$ is 
demonstrated and 
the characteristic of the LDT are clarified
on the basis of the one-parameter scaling theory together with the 
above hypothesis. The presence of critical subdiffusion, invariant nature
of critical perturbation strength and their dependency upon $M$ are fully
discussed. 
After these arguments, we reexamine the absence of LDT in 
the case of $M=2$ in Sec.\ref{sec:M=2}.
Finally, in Sec.\ref{sec:delocalized states}, characteristic of the normal 
diffusion in the delocalized states is discussed in some detail.
Summary and discussion are devoted in the last section.

\section{Models}
\label{sec:models}
We consider one-dimensional tightly binding disordered system represented by the lattece site
basis $|n\>$(n:integer) with the probability amplitude $\Psi_n$, which is driven by
time-dependent quasi-periodic perturbation.  The Schr\"{o}dinger equation
of the above system is represented by
\beq
\label{eq:model}
i \hbar \frac{\partial \Psi_n(t)}{\partial t}=\Psi_{n-1}(t)+\Psi_{n+1}(t)+V(n,t)\Psi_n(t), 
\eeq
where $V(n,t)$ is the time dependent on-site potential.
We deal with the following two cases, $V_A(n,t)$ and $V_B(n,t)$, as $V(n,t)$
with coherent periodic perturbation $f_\eps(t)$:
\beq
V(n,t)=
\begin{cases}
V_A(n,t)=V(n)[1+f_\eps(t)] & ({\rm A-model})\\
V_B(n,t)=V(n)f_\eps(t) & ({\rm B-model}).
\end{cases}
\eeq
The  coherent periodic perturbation $f_\eps(t)$ is given as,
\beq
    f_\eps(t)=\frac{\eps}{\sqrt{M}} \sum_i^M\cos(\omega_i t + \theta_i), 
\eeq
where $M$ and $\eps$ are number of the frequency component and 
the relative strength of the perturbation, respectively.
Note that the long-time average of the total power of the perturbation is normalized to 
$\overline{f_\eps(t)^2}=\eps^2/2$.
The frequencies $\{ \omega_i\}(i=1,...,M)$ are taken as mutually incommensurate numbers 
of order $O(1)$ given in Appendix \ref{app:omega}.
Here we take $\theta_i=0$($i=1,2,...M$)
to see long-term results that do not depend on the details 
of initial phases $\{\theta_i \}$.
The static on-site disorder potential is represented as $V(n)=Wv_n$.
$W$ denotes the strength of potential, and $v_n$  is 
uniform random variable with the range $[-1, 1]$ which is decorrelatd
between different sites. 
In the A-model, it becomes the Anderson model if we take $\eps=0$,
and the Anderson localization occurs. How the localization may become delocalized by 
increasing the perturbation strength $\eps$ is the main problem to be clarified.
On the other hand, the B-model is controlled by the combined parameter $\eps W$,
and if we take $\eps W=0$, the eigenstates are the Bloch states. The issue is how 
the ballistic motion of $\eps=0$ may make transition to a stochastic motion such 
as the normal diffusion by increasing $\eps$, which models stochastization of 
ballistic electrons by dynamical impurities.

We remark that time-dependent model (\ref{eq:model}) has an autonomous representation.
The isolated harmonic modes form a $M-$dimensional ladder of the eigenstate $|\{n_i\}\>$ 
which is assigned by the set of integers $\{n_i\}~(1\leq i\leq M)$ as
the quantum numbers and has the energy $E_h(\{n_i\}):=\sum_{i=1}^M\omega_in_i$. 
If we denote the eigenstate of 1D Anderson model of $\eps=0$ by $|N\>$, which are the 
Anderson localized state (A-model) or Bloch states (B-model) having the energy eigenvalue $E_N$,  
then Eq. (\ref{eq:model}) is equivalent to the autonomous Schr\"{o}dinger equation
describing the transition process among $(M+1)-$dimensional 
lattice of sites assigned by $(N,\{n_i\})$: let the probability amplitude of 
the quantum state $|N,\{n_i\}\>=|N\>|\{n_i\}\>$ be $\Phi(N,\{n_i\})$, then the 
Schr\"{o}dinger equation is represented by
\beq
\label{eq:modeld}
 i\hbar \frac{d\Phi(N,\{n_i\})}{dt}= \left[E_N+E_h(\{n_i\})\right] \Phi(N,\{n_i\}) \nn \\
+\frac{\eps}{\sqrt{M}}\sum_{N'}\sum_{j=1}^M W_{NN'}\Phi(N',n_1,.., n_j\pm 1,n_{j+1}...n_M), 
\eeq
where $W_{NN'}$ is the transition element $W\sum_n\<N|n\>v_n\<n|N'\>$ and $\{|n\>\}$ is 
 an orthonormalized basis set representing the  lattice site $n$.
The equivalent of Eq.(\ref{eq:modeld}) to the autonomous version of Eq(\ref{eq:model})
is presented in Appendix \ref{app:auto}.

We basically limit the perturbation strength to $\eps<0.3$, since
we are interested in how small $\eps$ may destroy the localization effect.
As $\eps$ increases far beyond the perturbation regime, the A-model will 
gradually approach to the B-model.

As the tool of  numerical integration of Eq.(\ref{eq:model}), 
we use the second-order symplectic integrator
\beq
\label{eq:YMY}
 U(\ell) = \e^{-i \Delta t \cos(n)/2\hbar}\e^{-iV(n,\ell\Delta t)/\hbar}\e^{-i \Delta t \cos(n)/2\hbar}
\eeq
with the small-enough time step $\Delta t=0.02 \sim 0.05$, where the value of Planck constant is
taken $\hbar=1/8$.
The system and ensemble sizes are $2^{15}-2^{16}$ and $10-50$, 
respectively, throughout this paper.
We use a localized state at $n=n_0$ 
 as the initial state and numerically observe the spread 
of the wavepacket measured by the mean square displacement (MSD), 
\beq
m_2(t) = \sum_{n}(n-n_0)^2 \< |\Psi(n,t)|^2 \>.
\eeq

In the limit $M\to\infty$, the  quasiperiodic perturbation $f_\eps(t)$ can be identified with the 
delta-correlated stochastic force $n(t)$ characterized by 
 $\<n(t)n(t^{`})\>=\eps_s^2 \delta(t-t^{`})$ with the strength $\eps_s$.
 In this paper, corresponding to A-model and B-model, 
 we consider the stochastic version of the two models in which the harmonic 
 force $f_\eps (t)$ is replaced by the noise force $n(t)=\eps_sn_1(t)$, 
 which varies at random in time uniformly in the range $[-1, 1]$:
\beq
 \begin{cases}
V_{SA}(n,t)=V(n)[1+\eps_s n_1(t)] & ({\rm SA-model}) \\ 
V_{SB}(n,t)=V(n)\eps_s n_1(t) & ({\rm SB-model}),
\end{cases}
\eeq
We call these  SA-model and SB-model, respectively.
In the SA-model, 
the localization is destroyed by the stochastic perturbation and 
the normal diffusion $m_2(t)=Dt$ with the diffusion constant
$D$ appears for $t\to \infty$ \cite{yamada98,yamada99}, 
as was first pointed out by Haken and 
his coworkers \cite{haken72,palenberg00}.
They predicted analytically the diffusion constant $D$ for the white Gaussian noise as
\beq
D=\lim_{t \to \infty}\dfrac{m_2(t)}{t} \propto \frac{\eps_s^2}{\eps_s^4+W^2/3}.
\label{eq:D-stochastic}
\eeq
for weak enough $\eps_s$. The diffusion constant increases as $D \propto \eps_s^2$ for  $\eps_s<<1$
and it reaches maximum at $\eps_s^*=\frac{W}{\sqrt{3}}$, 
and it finally decreases as $D \propto \eps_s^{-2}$.
The noise-induced diffusion has been extended for a random lattice driven by
the colored noise, including the hopping disorder effect \cite{moix13,knap17}.

For finite $M$, $f_\eps(t)$ can no longer be replaced by the random noise, and it playes as a
coherent dynamical perturbation, and the system corresponds to a quantum dynamical system
with $(M+1)$-degrees of freedom.

\section{Localized states of A-model}
\label{sec:localization}
First of all we show in this section the localization 
characteristics exhibited by our model Eq.(\ref{eq:model}).
The cases of $M=1,2$ are particularly focused on, and
a basic hypothesis to interpret all our numerical results is
discussed in connection with the localization characteristics
of our system.

\subsection{dynamics toward localization; localizing evolution}

\begin{figure}[htbp]
\begin{center}
\includegraphics[width=8.5cm]{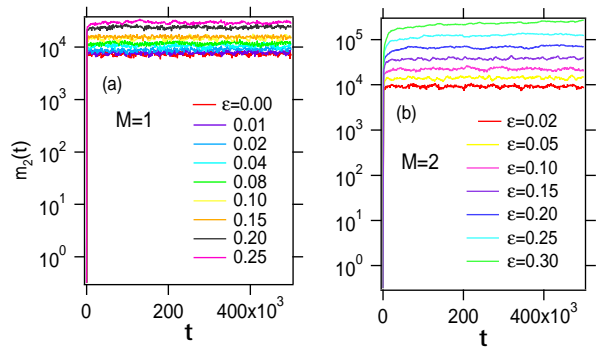}
\caption{\label{fig:c2-msd}(Color online)
The plots of $m_2(t)$ as a function of time
for different values of  $\eps$ increasing from bottom  to top
in the perturbed Anderson model.
 (a)$M=1$, $W=1.0$. 
 (b)$M=2$, $W=1.0$. 
Note that the horizontal axes are in the logarithmic scale.
}
\end{center}
\end{figure}

 Figure \ref{fig:c2-msd}(a) shows the time-dependence of MSD for
some typical cases of the monochromatically perturbed A-model,
for which the growth of time-dependence is saturated at a 
certain level. The the spread of the
wavepacket becomes larger as the perturbation
strength increases. This is the same tendency as
was observed for the Anderson map. In this paper, we
directly compute the localization length (LL) by
\beq
 \xi_M=\sqrt{m_{2}(\infty)}, 
\eeq
where $m_2(\infty)$ indicates the  numerically saturated MSD 
reached after a sufficiently long time evolution. 
For $M=1$ the localization is manifest.
Even in the case of $M=2$, localization occurs and the LL
increases as the perturbation strength increases
, as can be seen from the Fig.\ref{fig:c2-msd}(b).

Application of harmonic perturbation in general enhances the LL.
The enhancement of LL is conspicuous for $M=2$, and
the numerical evaluation of $\xi_M$ directly from the 
long time behavior of MSD
is possible only  in the limited range of $\eps < 0.4$.

\subsection{$W-$dependence of localization length}
\begin{figure}[htbp]
\begin{center}
\includegraphics[width=6.0cm]{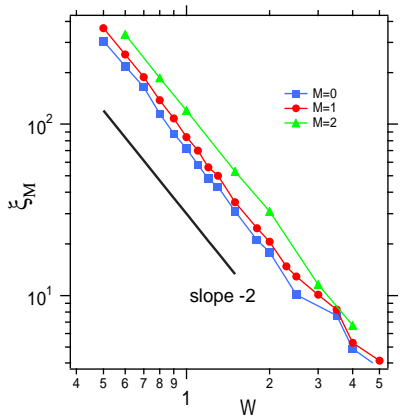}
\caption{\label{fig:c0c1-LL}(Color online)
Localization length  $\xi_M$ of the A-model
as a function of disorder strength $W$ for $M=0,1,2$ and $\eps=0.05$
}
\end{center}
\end{figure}

Figure \ref{fig:c0c1-LL} shows $W-$dependence of the LL
$\xi_M$ for $M = 0, 1, 2$. In all cases, it is naturally
found that for $\eps <<1$ the larger $W$, the stronger the
localization is, and the LL follows the rule
\beq
 \xi_M \sim \frac{A_M(\eps)}{W^2},
\eeq
where $A_M (\eps)$ depends on $M$ and $\eps$. 
The $W^{-2}-$dependence of the LL has been commonly observed in the case 
of quantum map systems \cite{yamada20}.
For $M=1$ the persistence of  localization can be expected as is argued
 in  Appendix \ref{app:M=1dash}.

\subsection{$\eps-$dependence of localization length ($\eps<<1$)}
\begin{figure}[htbp]
\begin{center}
\includegraphics[width=9.4cm]{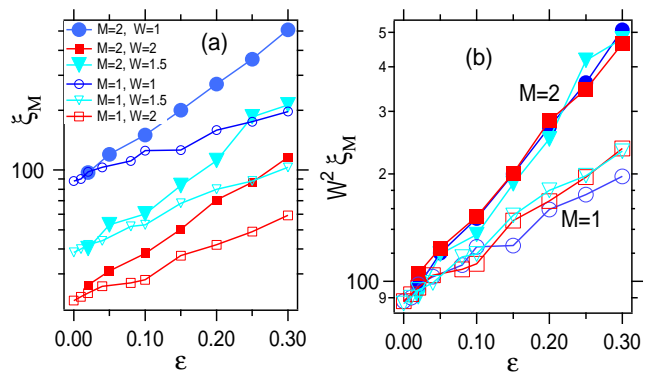}
\caption{\label{fig:LL-M1M2-scaled}(Color online)
(a)Localization length  $\xi_M$ of the A-model
as a function of perturbation strength $\eps$ for $M=1,2$ and 
$W=1.0, 1.5, 2.0$. 
(b) $\xi_M W^2$ as a function of $\eps$.
Note that the vertical axes are logarithmic scale.
}
\end{center}
\end{figure}

Figure  \ref{fig:LL-M1M2-scaled}(a) shows the result 
of the $\eps-$dependence in
the A-model of $M = 1,M = 2$ for some $W$'s. It is obvi-
ous that the LL grows exponentially as the perturbation
strength $\eps$ increases in the all cases:
\beq
\xi_M \sim \e^{c_M\eps}.
\eeq
When $W$ is the same, the exponentially growth rate $c_M$ of 
$M=2$ is larger
than that of $M=1$, and it can be seen that the coefficient
$c_M$ does not depend on the disorder strength $W$. 
To confirm this more concretely, we plot the $\eps$-dependence
in the  Fig.\ref{fig:LL-M1M2-scaled}(b) of the scaled LL $\xi \times W^2$. 
At least when $\eps$ is small ($\eps < 0.3$), they all overlap 
well, and the coefficient $c_M$ is almost
constant and has no $W-$dependence. Therefore,\\
\beq
 \xi_M \simeq \frac{\exp\{c_M\eps\}}{W^2}.
\eeq
This is similar to what was found  for the monochromatically 
perturbed Anderson map \cite{yamada18} in a small region of $\eps$.

Although it is difficult to obtain the LL $\xi_M$ 
directly from the long time behavior of MSD, it can be expected that a
similar tendency to the cases of M = 1 and M = 2 will
be observed even in the localized region of $M\geq 3$ 
for small enough $\eps$.
However, as is the case in the high-dimensional 
disordered lattices and also in the Anderson map system,
if localization-delocalization transition (LDT)
takes place at some critical $\eps_c$, the LL grows 
divergently as $\eps\to\eps_c$

\subsection{$\eps-$dependence of localization length for large $\eps$}
\begin{figure}[htbp]
\begin{center}
\includegraphics[width=4.6cm]{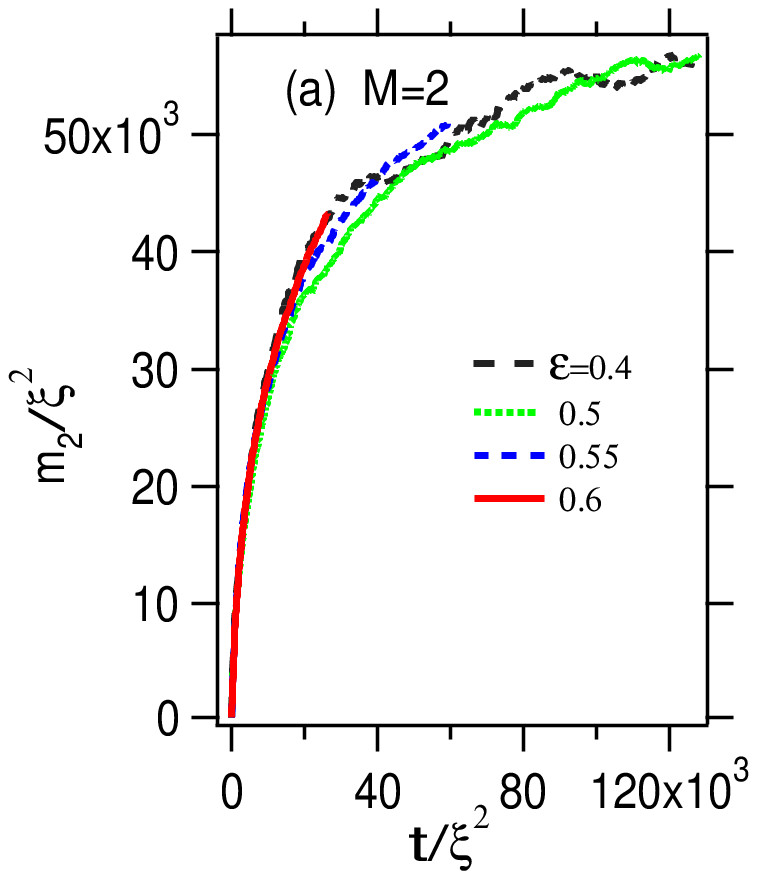}
\hspace{-8mm}
\includegraphics[width=4.6cm]{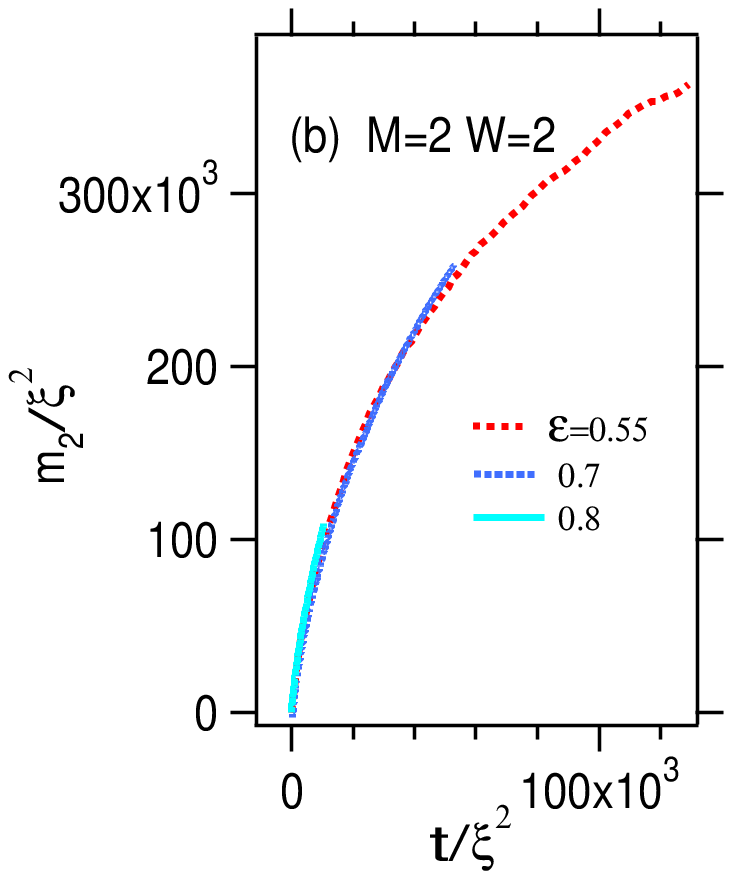}
\caption{\label{fig:c2-msd-ab-scale}(Color online)
Scaling peoperty 
$m_2(t)/\xi(\eps)^2$ as a function of $t/\xi(\eps)^2$
in the dichromatically perturbed A-model of $W=1$  for 
various $\eps$'s.
(a)$\eps=0.40,0.50,0.55,0.60$ and (b)$\eps=0.55,0.70,0.80$.
}
\end{center}
\end{figure}

We observed that, at least, the wavepacket localizes
completely when $M=2$ in the region where the perturbation strength 
is relatively small $\eps < 0.4$.
We would like to investigate the localization length $\xi_M$
for $M=1$ and $M=2$ when $\eps$ increases beyond the perturbation region. 
In the region where $\eps$ is large, 
the localization length $\xi_M$ cannot be estimated directly by the 
saturation level of the MSD.\\ 
Here, we try to determine $\xi_M$
 indirectly by supposing that the MSD data follows the common
scaling from independent of $\eps$ as
\beq
m_2(t)\sim \xi(\eps)^2 F\left(\frac{t}{\xi(\eps)^2}\right), 
\label{eq:M1scaling}
\eeq
where $F(x)$ is a scaling function. 
To confirm this, we show in Fig.\ref{fig:c2-msd-ab-scale}
the plots of $m_2/\xi(\eps)^2$ as a function of $t/\xi(\eps)^2$,
which manifests the scaling hypothesis of Eq.(\ref{eq:M1scaling}).

We can estimate the localization length $\xi_M(\eps)$ 
by using, and sometimes by repeatedly using, the scaling 
hypothesis Eq.(\ref{eq:M1scaling}) even for $\eps>0.4$.

\begin{figure}[htbp]
\begin{center}
\includegraphics[width=6.5cm]{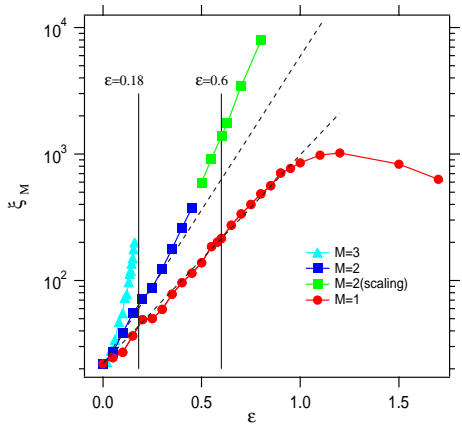}
\caption{\label{fig:loc-len-eps}(Color online)
Localization length as a function of $\eps$ for
$M=1,2,3$ with $W=1$.
Some LL of $M=2$ are obtained by the scaling hypothesis
Eq.(\ref{eq:M1scaling}) for $\eps >0.4 $.
Note that the horizontal axis is in logarithmic scale.
The dashed lines are $e^{5.5\eps}$ and $e^{3.8\eps}$, respectively.
The lines $\eps=0.18$ and $\eps=0.6$ are shown as a reference.
}
\end{center}
\end{figure}

Figure \ref{fig:loc-len-eps}
 shows the $\eps-$dependence of a wide range of
the localization lengths, including indirectly determined
$\xi_M$ with the scaling hypothesis (\ref{eq:M1scaling}). 
For comparison, $\xi_M$ of $M=3$, which exhibits a clear 
DLT as discussed in detail later, is also shown.
The localization, of course, occurs in the case of $M=1$.

Then what is the difference of the localizations between
the case of $M=1$
and the cases of $M=2$. 
In both cases of $M=1$ and $M=2$,
the localization length grow exponentially when the $\eps$ is
small enough ($\eps<0.8$ for $M=1$ and $\eps< 0.3$ for $M=2$).

For $M=1$, it is obvious that the localization occurs no matter how
large $\eps$ may be, but, as for $M=2$, the presence or absence of 
DLT is still unclear.
We will discuss again the persistence of localization for $M=2$ 
in Sect. \ref{sec:M=2} after the next Sect.\ref{sec:LDT} 
in which the presence of DLT 
is confirmed for $M\geq 3$.
In the next subsection we consider 
the {\it substantial dimension} of our system which may dominate
the upperbound dimension of localization.

\subsection{The effective dimension}

Our model (\ref{eq:model}) is very similar to that of the AM 
perturbed by $M$ harmonic modes, which
is represented by the symplectic propagator (\ref{eq:YMY}) of $\Delta t=1$
It is formally transformed into $d(=M+1)-$dimensional quasi-random lattice
by the so called Maryland transformation \cite{yamada20}, and $d=2$, i.e. $M=1$, 
is the upper-bound of dimension in which deloclization does not happen.
Unlike this, in the present model the numerical observations suggest $M=2$ 
may be the upper-bound dimension of the localization. 
Why is there such a difference is? 

In the case of AM, time is not continuous and there is no conserved quantity.
However, in the present case, Eq.(\ref{eq:model}) is rewritten as Eq.(\ref{eq:auto3})
 given in Appendix \ref{app:auto}  which yields a severe constraint of energy conservation.
In the transition process by the interaction among the harmonic modes and 
the isolated 1D random lattice the constraint due to the energy conservation
\beq
  \left|\sum_{m=1}^{M} \Delta n_m\omega_m\right| <\frac{|E_N-E_{N'}|}{\hbar} < \frac{C}{\hbar}
\label{eq:bound}      
\eeq
exists, where $E_N$ and $E_{N'}$ are the energies of the localized eigenstates 
and $\Delta n_i=n_i'-n_i$ is the change of excitation number of $i$-th harmonic mode.  
The upper-bound of $C={\rm Max}\{|E_N-E_{N'}|\}$ is 
estimated as $C<4+2W$. IF $C=0$, the number of degrees of freedom exactly
reduces by exactly 1,  and
\beq
   d_f=(M-1)+1=M
\label{eq:new-d}
\eeq 
is the effective dimension of the system.
However, since $C$ is finite, the system should be
regarded as the ``quasi-$d_f$'' dimensional system
in the sense that $M-1$ quantum numbers can 
arbitrarily be changed but the $M$-th mode is
restricted by Eq.(\ref{eq:bound}). 
If $d_f$ corresponds to the spatial
dimension of the irregular lattice, then the maximal dimension in which
only the localization exist can be $d_f=M=2$
if the scaling theory of the localization is followed.
In the present paper
we interpret the results presented below on the hypothesis that 
Eq.(\ref{eq:new-d}) is the ``effective dimension''.  
We emphasize that the hypothesis was not taken into account in our previous letter,
and $M+1$, instead of $M=d_f$, was used as the system dimension \cite{yamada21}.

\section{Localization-Delocalization transition: A-model}
\label{sec:LDT}
In this section, we investigate LDT of the A-model with increasing the number of colors
from $M=3$ to $M=7$ while paying attention to the correspondence with 
result in the Anderson map system.
The case of $M=2$, which has a large localization length 
but is expected to have no LDT, will be also rediscussed 
in next section.

\subsection{dynamical LDT}
\begin{figure*}[t]
\begin{center}
\includegraphics[width=12.0cm]{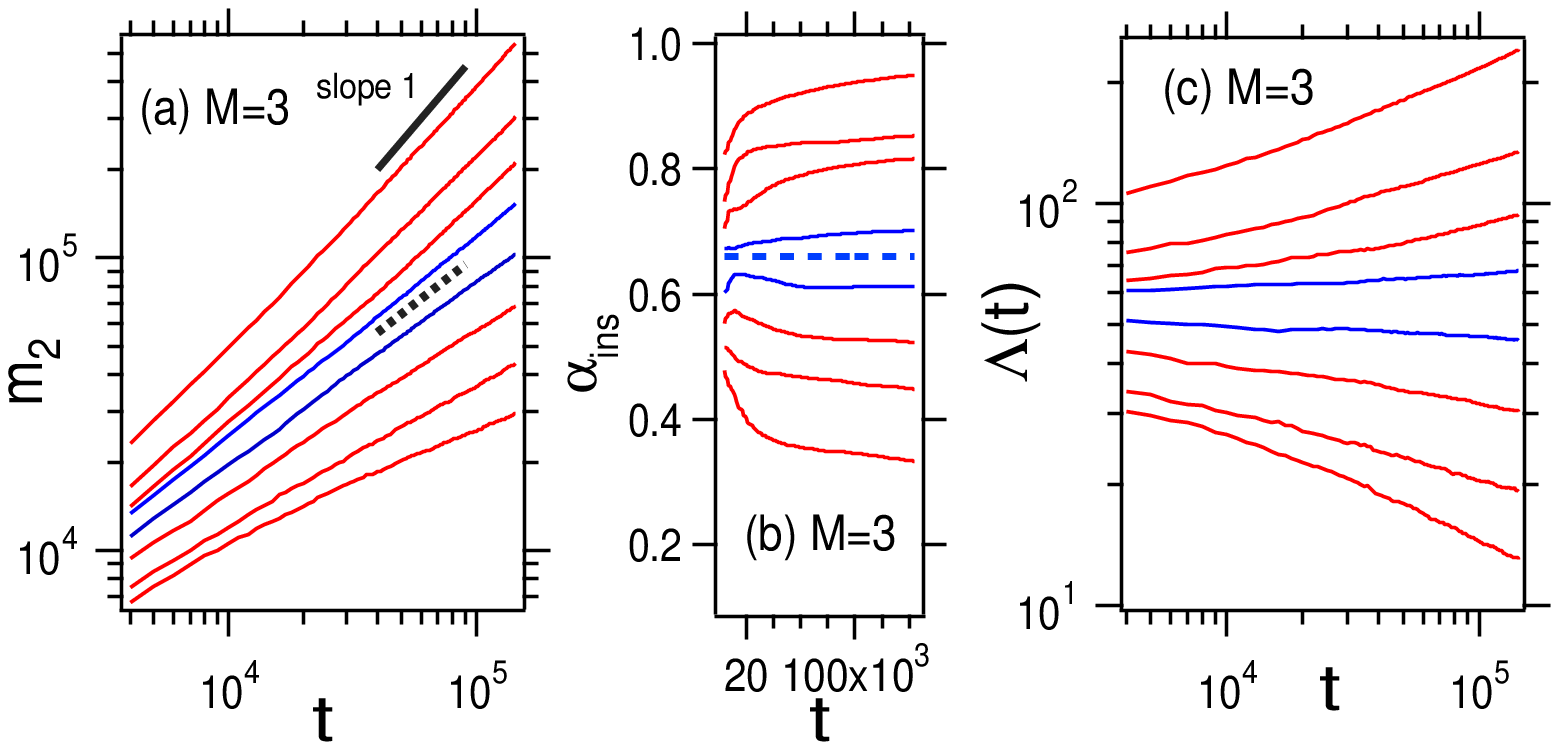}
\hspace{3.0mm}
\includegraphics[width=12.0cm]{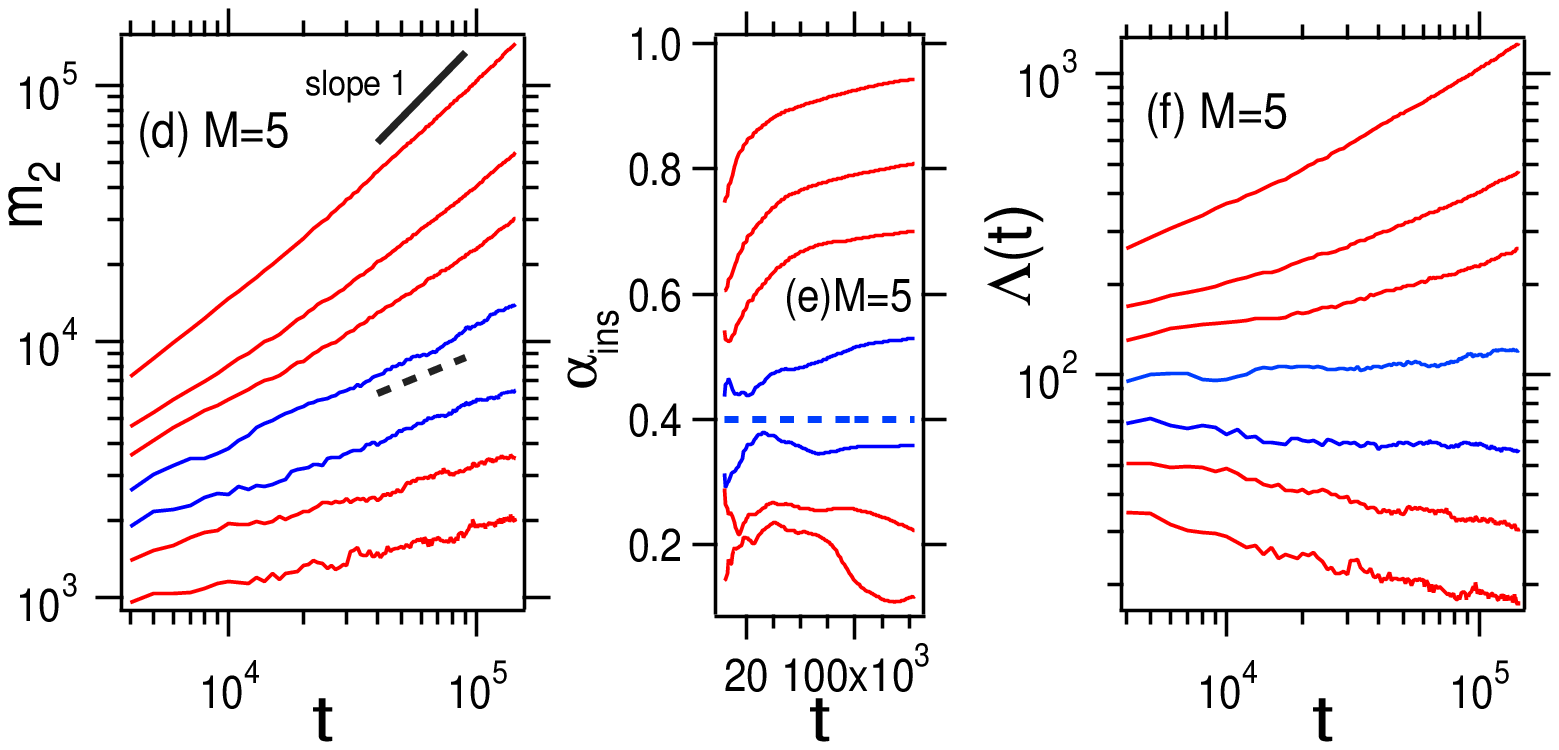}
\caption{\label{fig:c3-s1s3-msd}(Color online) 
Localization-delocalization transition for the A-model
 exhibited by the change of time-dependence of 
MSD: (a),(b), and (c) for $M=3$ and $W=1$ and (d),(e) and 
(f) for $M=5$ and the same $W=1$. 
(a) The double-logarithmic plots of MSD $m_2(t)$, 
(b) the diffusion index $\alpha_{ins}(t)$
and (c) the scaled MSD $\Lambda(\eps,t)=m_2(t)/t^{\alpha_c}$, where $\alpha_c=0.66$ 
as functions of time for increasing perturbation strengths 
$\eps=0.17,0.18,0.19,0.20,0.21,0.22,0.23,0.25$ from below.
The broken line in (b) indicates the critical subdiffusion line
$\alpha_{ins}(t)=\alpha_c=0.66$ predicted by the scaling theory.
(d)(e) and (f) are the counterparts of (a)(b) and (c), respectively,
for $M=5$, where $\eps$ is increased as $\eps=0.05,0.06,0.07,0.08,0.09,0.10,0.12$
from below and $\alpha_c=0.40$.
} 
\end{center}
\end{figure*}

In Fig.\ref{fig:c3-s1s3-msd},  typical examples indicating the LDT
for $M\geq 3$ are depicted. They are the double logarithmic plot of the time 
evolution of MSD for an increasing series of the perturbation strength $\eps$.
For both examples one can recognize that with an increase in $\eps$ the 
time evolution of MSD exhibits a transition from a saturating behavior 
to a straight line of slope 1 implying the normal diffusion $m_2\propto t$. 
A remarkable fact is that the transition proceeds through a time evolution
represented by a straight increase with a fractional slope $0<\alpha<1$ at a 
particular value $\eps=\eps_c$. It can be regarded as the critical subdiffusion 
$m_2\propto t^\alpha$. Indeed, for $M\geq 3$ the numerical results
indicate that the asymptotic behavior of the MSD in the limit $t\to\infty$ changes as
\beq
\label{eq:transition}
m_2(t) \sim 
\begin{cases}
t^0  {\rm (localization)} & \eps<\eps_c  \\
t^{\alpha}  {\rm (subdiffusion)} & \eps \simeq \eps_c  \\  
t^{1}  {\rm (delocalization)} &\eps>\eps_c
\end{cases}
\eeq
which fully follows the numerical observations
in AM and SM \cite{yamada20}.

To confirm numerically the critical behavior represented by Eq.(\ref{eq:transition}),
it is very convenient to introduce the local diffusion exponent
defined as the instantaneous slope of the log-log plot of MSD
\beq
   \alpha_{ins}(t)=\dfrac{d\log m_2(t)}{d\log t}.
\eeq
as a function of $t$, where $m_2(t)$ is appropriately smoothed.

Figure \ref{fig:c3-s1s3-msd}(a)-(c) and (d)-(f) are examples of the
transition process modeled by Eq.(\ref{eq:transition}) for $M=3$ and
$M=5$, respectively. (a) and (d) represent the change of MSD from the
localized states to the normal diffusion state. Transition from the localized
state to the normal diffusion is directly recognized by the change of $\alpha_{ins}(t)$ plots
 demonstrated in (b) and (e). It either decays to 0 or 
increases toward 1, and  it keeps a constant value only at a particular $\eps=\eps_c$, 
indicated by broken lines, which means the existence of the critical
subdiffusion $m_2(t) \propto t^{\alpha_{c}}$ at $\eps=\eps_c$, 
where $0.60<\alpha_c<0.70$ and $0.35<\alpha_c<0.45$ in (b) and (e), respectively.

These facts suggest the so called one-parameter scaling theory,
 which was successfully used in the analyses of AM and SM, is applicable
to our model, identifying the effective dimension Eq.(\ref{eq:new-d}) as the
dimension $d$ of the random system. It predicts the critical subdiffusion 
index as
\beq
\label{eq:alpha-M}
  \alpha=\frac{2}{d_f}=\frac{2}{M}. 
\eeq
The theoretical value $\alpha\sim 0.66$ for $M=3$ and $\alpha\sim 0.40$ 
for $M=5$ are drawn in (b) and (e) by broken lines, respectively. 
Agreement with the critical lines suggested by $\alpha_{ins}(t)$ plots is evident. 
We note that in our preliminary report we took $d_f=M+1$, instead of Eq.(\ref{eq:new-d}), 
because the restriction (\ref{eq:bound}) was not taken into account. 
However, as $M$ increases beyond 5, Eq.(\ref{eq:alpha-M}) become less cofirmative.

To make a further check of the LDT close to the critical point, 
it is instructive to use the MSD $\Lambda(t)$ divided by the 
critical subdiffusive increase:
\beq
\label{eq:Lambda}
\Lambda(t)\equiv \dfrac{m_2(t)}{t^{\frac{2}{M}}}.
\eeq
Then $\Lambda(t) \simeq const.$ indicates the critical point, and 
$\Lambda(t)$ grows upward for $\eps>\eps_c$, while it decays downward for 
$\eps<\eps_c$, as are seen in Fig.\ref{fig:c3-s1s3-msd}(c) and (f).
The feature that the $\Lambda(t)$ curves expands to form a trampet-like pattern
suggests the existence of the LDT.

\begin{figure}[htbp]
\begin{center}
\includegraphics[width=6.0cm]{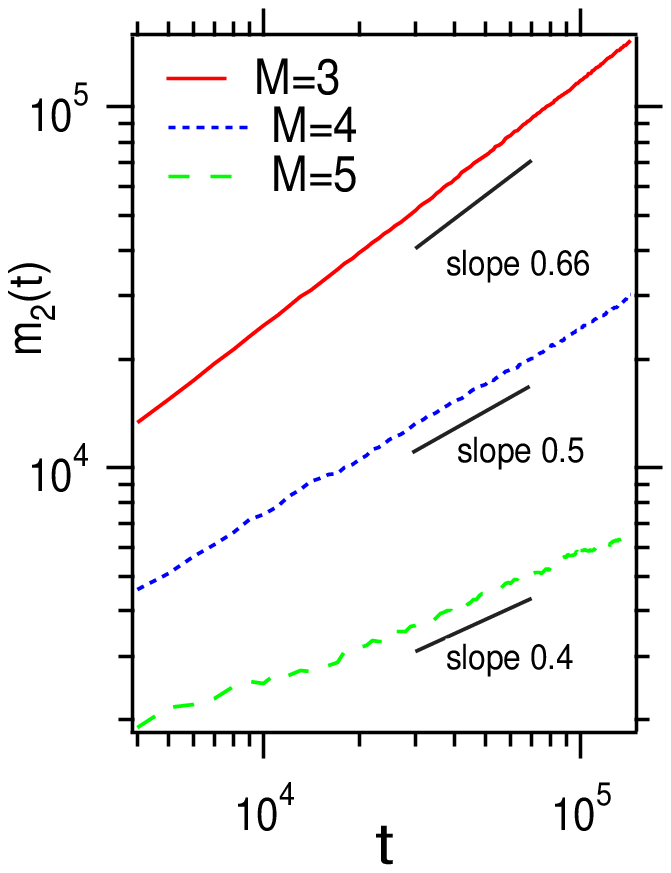}
\caption{\label{fig:M3M4M5-epsc}(Color online)
The double-logarithmic plots of $m_2(t)$ 
as a function of time near the critical pints $\eps_c$ in
the polychromatically perturbed 1D Anderson model ($M=3,4,5$ from top) with $W=1$.
}
\end{center}
\end{figure}

As are shown in Fig.\ref{fig:M3M4M5-epsc}, we confirm that the critical sub-
diffusion can be observed at certain critical point 
$\eps=\eps_c$ even if $M$ is increased beyond 3, and it 
is evident that the subdiffusion index $\alpha$ at the critical 
point decreases as $M$ increases, and it is numerically 
consistent with the prediction of Eq.(\ref{eq:alpha-M}).

\subsection{$M-$dependence of the scaling property for the LDT}
\begin{figure}[htbp]
\begin{center}
\includegraphics[width=7.0cm]{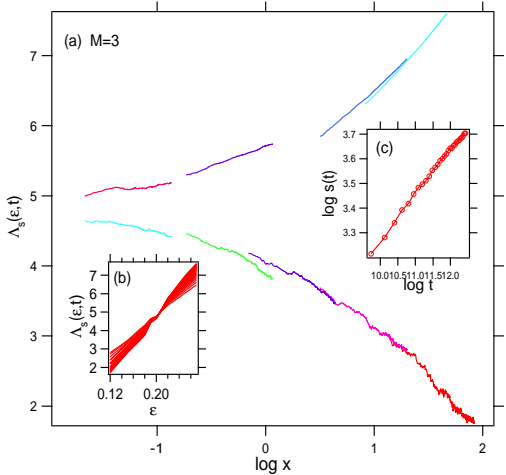}
\caption{\label{fig:c3-nu}(Color online)
The results of the critical scaling analysis for 
trichromatically perturbed A-model ($M=3$) with $W=1.0$.
(a)The scaled MSD $\Lambda_s(\eps,t)=\log \Lambda(\eps,t)$ 
as a function of $x=\xi_M(\eps)t^{\alpha/2\nu}$ for some values 
of $\eps$.
(b)The scaled $\Lambda_s(\eps,t)$ with $\alpha=0.66$ 
as a function of $\eps$  for some pick up times.
The crossing point is $\eps_c \simeq 0.21$.
(c)$s(t)$ as a function of $t$.
The critical exponent $\nu \simeq 1.81$ is determined by a scaling relation 
Eq.(\ref{eq:s(t)}) by the least-square fit.
}
\end{center}
\end{figure}

In Fig.\ref{fig:c3-nu}(a), we show result of finit-time scaling 
analysys for the A-model of $M=3$.
The method used here is the same as that used in the paper \cite{yamada20}.
We choose the following quantity as a scaling variable
\beq
\Lambda_s(\eps,t)=\log \Lambda(\eps,t) =F(x),
\eeq
by shifting the time axis to $x$:
\beq
x=\xi_M(\eps)t^{\alpha/2\nu}, 
\eeq
for different values of $\eps$ by using  critical exponent $\nu$ to characterize 
the divergence of the localization length around the LDT:
\beq
 \xi_M \sim |\eps-\eps_c|^{-\nu}.
\eeq
$F(x)$ is a differentiable scaling function and $\alpha$ is the diffusion index.

Figure \ref{fig:c3-nu}(b) shows a plot of $\Lambda_s(t)$ 
as a function of $\eps$ at several times $t$, and 
it can be seen that this intersects at the critical point $\eps_c$.
In addition, Fig.\ref{fig:c3-nu}(c) shows a plot of 
\beq
s(t)=\frac{\Lambda_s(\eps,t)-\Lambda_s(\eps_c,t)}{|\eps_c-\eps|}  \propto  t^{\alpha/2\nu} 
\label{eq:s(t)}
\eeq
as a function of $t$, and the critical localization exponent $\nu$ 
is determined by best fitting this slope.
This is consistent with formation of the one-parameter scaling theory (OPST) of the localization.
As a result, even in the A-model, the OPST is well established for the LDT
regardless of the number of colors $M$ and the disorder strength $W$. 

The critical exponent evaluated using the data ($\alpha=0.66, \eps_c=0.21$) 
at $W=1$ for $M=3$ is $\nu \simeq 1.81$.
The same is true for the other $M (\geq 4)$ color perturbations.
Appendix \ref{app:critical-scaling-data} shows 
the results of the finite-time scaling analysis when $M=4$ and $M=7$.
These results are similar to that of  AM system perturbed by the $(M-1)$ colors
and of numerical calculations using finite-size scaling 
in the $d_f(=M)-$dimensional random systems.
Note that pursuing the numerical value of $\nu$ with high accuracy is not 
the purpose of this paper.

\subsection{$M-$dependence of critical strength $\eps_c$}

\begin{figure}[htbp]
\begin{center}
\includegraphics[width=4.1cm]{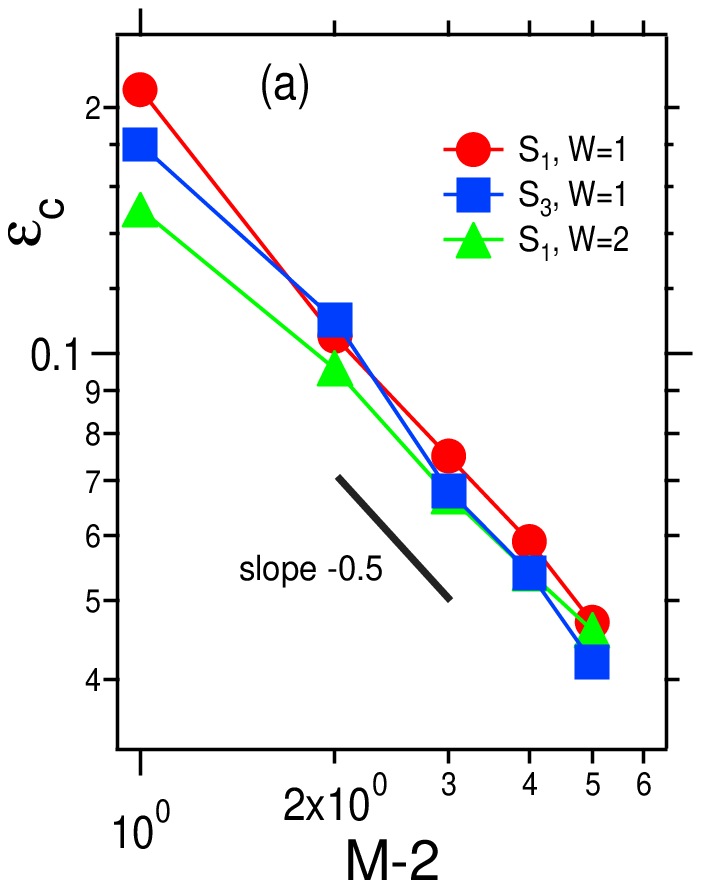}
\hspace{-3.0mm}
\includegraphics[width=4.6cm]{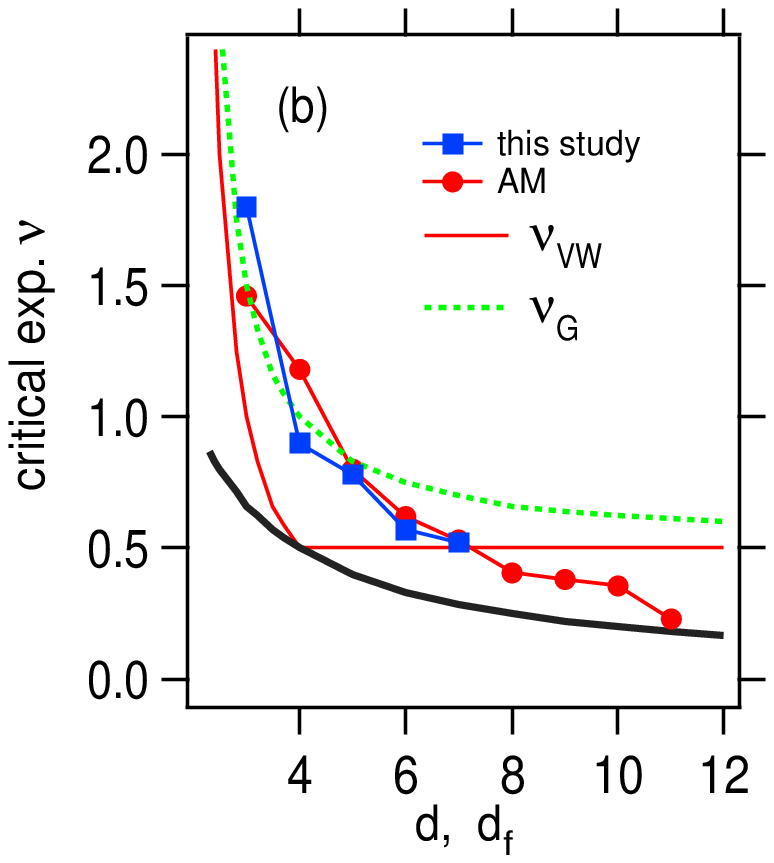}
\caption{\label{fig:eps_c-(M-1)}(Color online)
(a)The critical perturbation strength $\eps_c$ as a function of  $(M-2)$ 
for A-model with $W=1$. The black solid line shows $\eps_c \propto 1/(M-2)^{0.5}$.
(b)The effective dimensionality $d_f=M$ dependence 
of the critical exponent $\nu$
which characterizes the critical dynamics. 
The  red solid line and green dashed line are 
 the results of the analytical prediction by 
$\nu_{VW}$ and $\nu_{G}$, respectively.
Thick line denotes the lower bound by the Harris' critical inequality.
}
\end{center}
\end{figure}

Return to the story of critical perturbation strength $\eps_c$.
As is seen in Fig.\ref{fig:eps_c-(M-1)}(a),  $\eps_c$ definitely decreases with increase
in $M$ for $M\geq 3$.
Looking upon $\eps_c$ as the function of $M-2$, the double-logarithmic plots are on a straight 
line with the approximate tangent $-0.5$, namely
\beq
\eps_c \sim \frac{1}{(M-2)^\delta},~~ \delta \simeq 0.5,
\label{eq:epsc-M}
\eeq
This result suggests that $\eps_c$ diverges at $M=2$, and the LDL transition do not
exists at $M=2$.
Evident dependence of $\eps_c$ on $M$ for large $M$ contradict with 
the prediction of the SCT \cite{yamada20}.

The critical exponent $\nu$, which characterizes the divergence of 
the localization length at the critical point 
 is numerically evaluated, and 
 plotted against $M$, as shown in Fig.\ref{fig:eps_c-(M-1)}(b).
As a result, it can be seen that the tendency for $M\geq 3$ is close to 
that in the Anderson map.

\subsection{$W-$dependence of the critical point $\eps_c$}

\begin{figure}[htbp]
\begin{center}
\includegraphics[width=5.5cm]{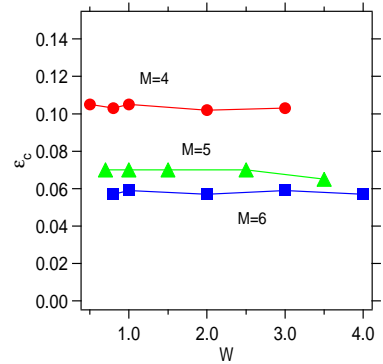}
\caption{\label{fig:epsc-W}(Color online)
The critical perturbation strength $\eps_c$ as a function of  $W$ 
in A-model of $M=4,5,6$ (from top).
}
\end{center}
\end{figure}

Figure \ref{fig:epsc-W} shows the $W-$dependence of the 
critical perturbation strength $\eps_c$ for $M=4$,$M=5$ and $M=6$.
From this result, it can be inferred that the critical perturbation strength $\eps_c$
of the LDT keeps an almost constant value insensitive to the disorder
strength $W$ and is only determined by the number of colors $M$.
Such a feature agrees with that observed in the Anderson map system with 
$M \geq 2$ for which the LDT emerges.

\begin{figure}[htbp]
\begin{center}
\includegraphics[width=4.2cm]{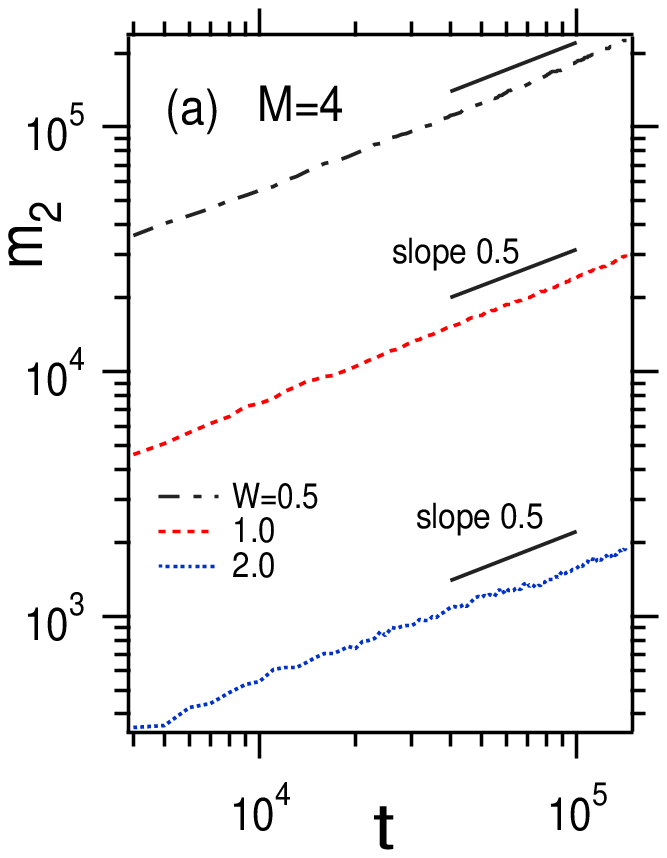}
\hspace{-2mm}
\includegraphics[width=4.2cm]{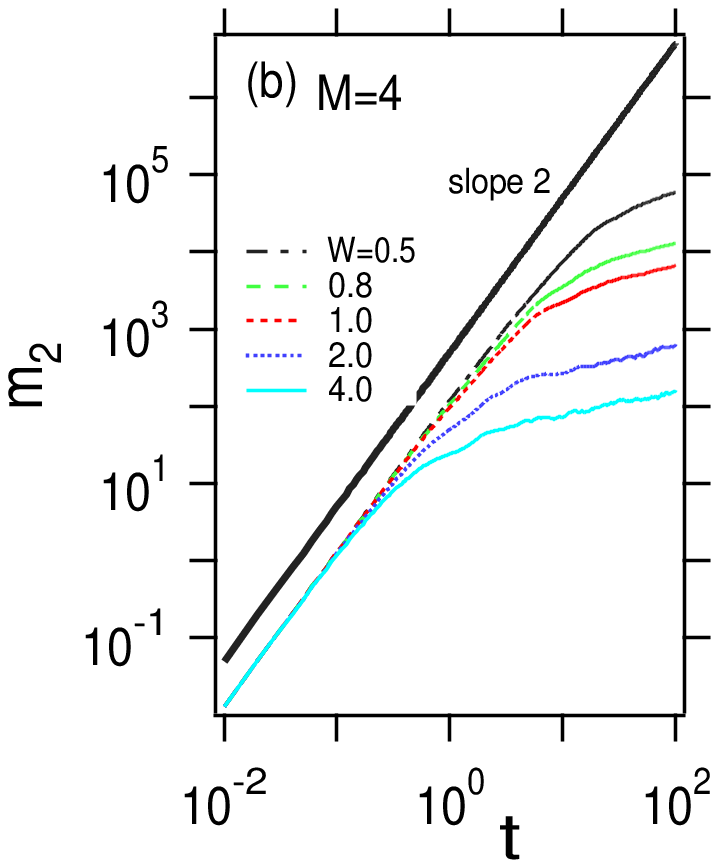}
\hspace{5mm}
\includegraphics[width=4.2cm]{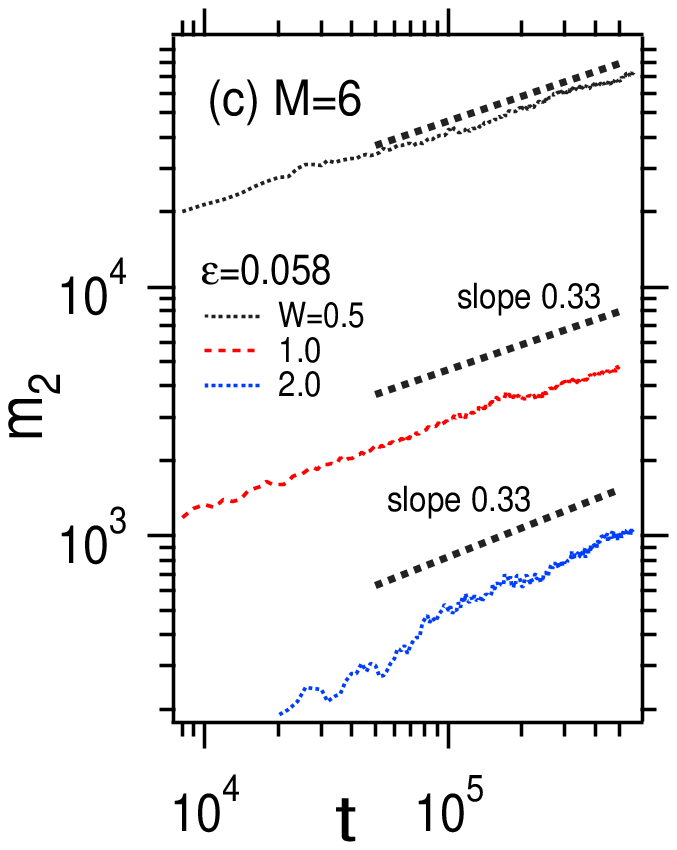}
\hspace{-2mm}
\includegraphics[width=4.2cm]{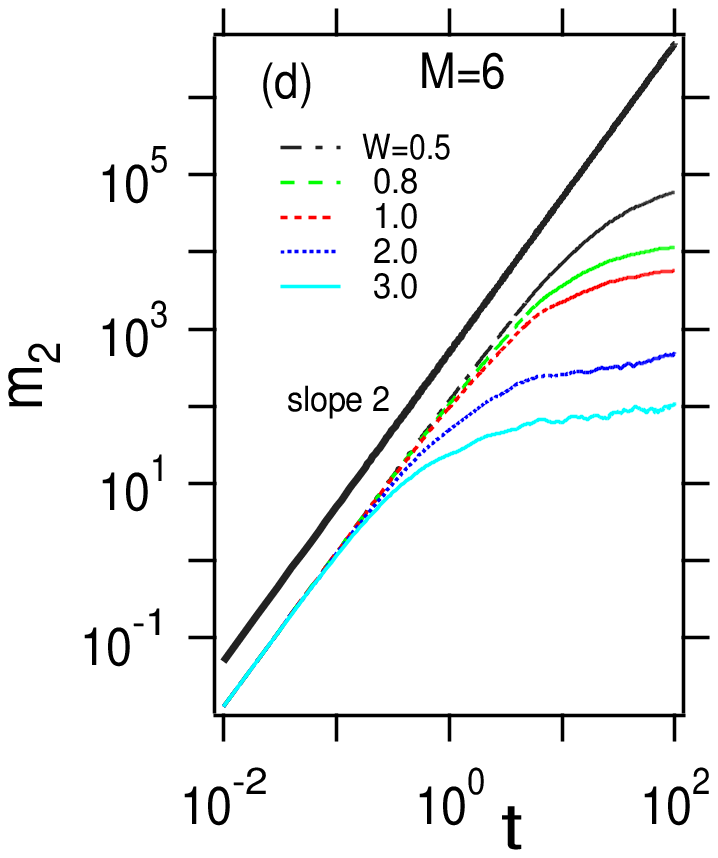}
\caption{\label{fig:c4c6-msd}(Color online)
The double-logarithmic plots of $m_2(t)$ 
as a function of time $t$ near the critical pints $\eps_c$  for different $W$ in
the A-model of (a)$M=4$ and of (c)$M=6$.
The panels (b) and (d) show the enlarged view of the short-time region $t<10^2$
in the double-logarithmic plots of $m_2(t)$ 
in the A-model of  (b)$M=4$ and (d)$M=6$.
}
\end{center}
\end{figure}

We show another direct evidences manifesting
that the magnitude of $W$ does not influence the LDT.
The time evolution of the MSD at $\eps_c$ is shown for several values of $W$
in Fig.\ref{fig:c4c6-msd}. First, looking at the case of $M=4$ in 
Fig. \ref{fig:c4c6-msd}(a) the spread  $m_2(t)$ of wavepacket becomes larger
with decrease in $W$, as is expected. But in all case we see that for the
same $\eps=\eps_c=0.115$ a subdiffisive increase at the same index $\alpha \simeq 0.5$ 
emerge regardless of $W$.
Similarly, in the case of $M=6$, regardless of $W$,
for the same $\eps_c \simeq 0.058$ the subdiffusion of $\alpha \simeq 0.33$
emerges, as are seen in Fig.\ref{fig:c4c6-msd}(c).

Figure \ref{fig:c4c6-msd}(b) and (d) are the enlargement of the 
initial growth of MSD for $t<10^2$ of (a) and (c), respectively.
In all cases, the wavepacket starts with a ballistic expansion
$m_2 \sim t^2$ and changes to exhibit the critical subdiffusion 
after a lapse of characteristic time. A paradoxical fact is that 
the characteristic time required for realizing the subdiffusive 
delocalization decreases with increase in the disorder strength $W$.

The larger the $W$, the stronger the localization, and
as the localization becomes stronger, delocalization occurs more promptly.
This means that what is important for delocalization is not to activate
the ballistic expansion of the wavepacket, but to promote its decomposition 
into particle-like quantum states called localized states due to 
the accumulation of scattering by disorder.
Delocalization emerges as the diffusive motion over the localized 
particle-like states.

\section{Reconsideration of weak dynamical localization for $M=2$}
\label{sec:M=2}
We return to the problem on the presence of DLT in the case of $M=2$.

\begin{figure*}[t]
\begin{center}
\includegraphics[width=6.0cm]{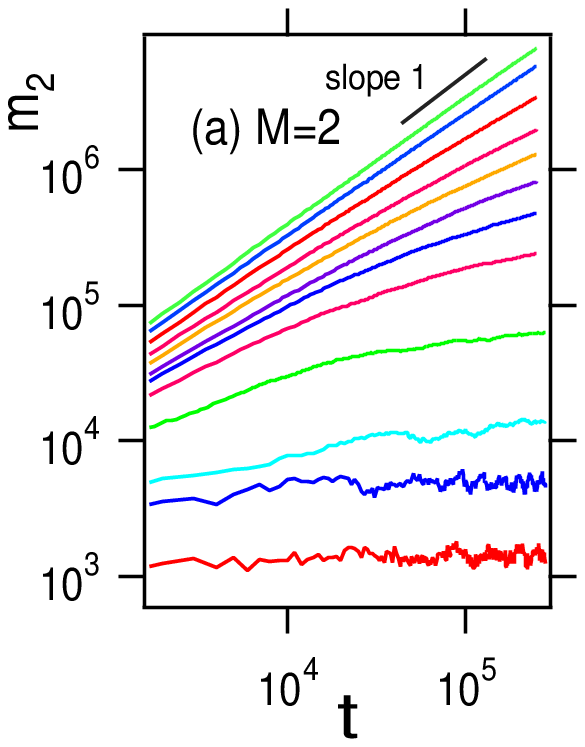}
\hspace{-10mm}
\includegraphics[width=4.2cm]{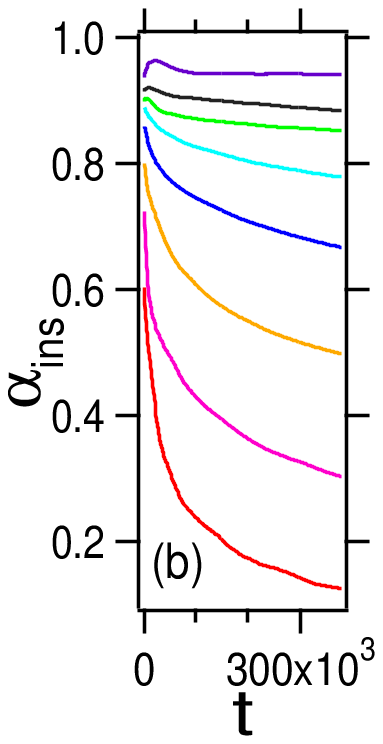}
\hspace{-10mm}
\includegraphics[width=6.0cm]{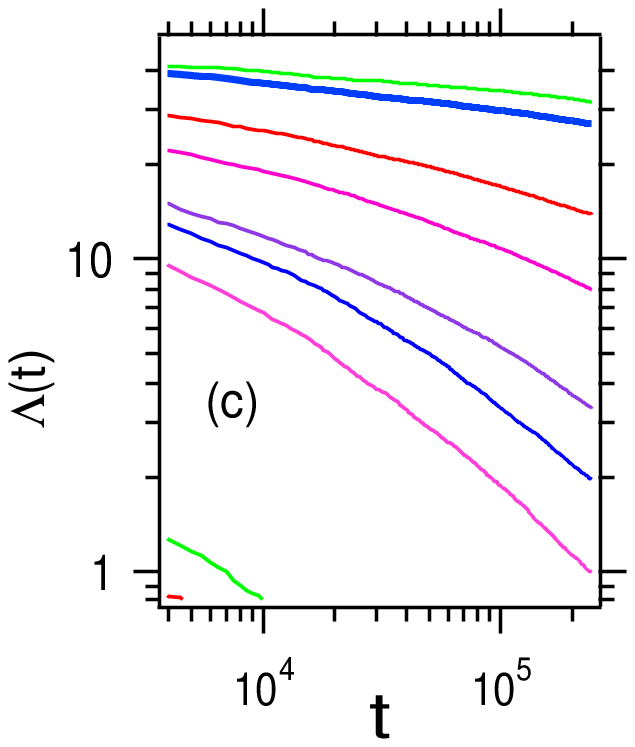}
\caption{\label{fig:M2-lambda} (Color online)
(a)The double-logarithmic plots of $m_2(t)$
as a function of time for some values of the perturbation strength $\eps$ 
 increasing from $\eps=0.1$ to $\eps=1.3$
in the A-model of $M=2$ with $W=1$.
(b)The instantaneous diffusion index $\alpha_{ins}(t)$ as a function of time.
 (c)The double-logarithmic plots of 
the scaled MSD
$\Lambda(\alpha=1,t)=\frac{m_2(t)}{t}$ as a function of time 
for some $\eps's$ from $\eps=0.1$ to $\eps=1.3$.
}
\end{center}
\end{figure*}

It is very hard to numerically prove the persistence of localization,
either by directly pursuing time evolution dynamics or by applying
the scaling hypothesis. (See  Fig.\ref{fig:M2-lambda}(a).)
However, there are some evidences manifesting that there exists no critical 
subdiffusion such that $m_2 \propto  t^\alpha$ with $0<\alpha<1$.
To numerically prove the presence of critical subdiffusion, an explicit
method is to use the $\alpha_{ins}(t)$ plots presented in the previous section. 
We examine in  Fig.\ref{fig:M2-lambda}(b) the $\alpha_{ins}(t)$ plots 
for $M=2$. All the curves go downward and it can hardly be expected 
that a horizontal line locates in the narrow gap between the 
line $\alpha=1$ and the uppermost downward curve, which implies that 
$\alpha=1$ plays the role of ``critical diffusion''.
This fact is consistent with the results of previous section represented
by Eq.(\ref{eq:alpha-M}) and Eq.(\ref{eq:epsc-M}) for $M\geq 3$, 
which predict $\alpha=1$ and $\eps_c=\infty$,
respectively, for $M=2$. \\
We further examine in Fig.\ref{fig:M2-lambda}(c), the $\Lambda$ plots Eq.(\ref{eq:Lambda}), namely 
the MSD scaled by the critical MSD $m_2(t)\propto t^{\alpha}$,
which is $\Lambda(\alpha=1,t)=\frac{m_2(t)}{t}$ supposing $\alpha=1$. 
All the curves go downward for $t>>1$
to form the lower-half of the pre-critical trampet pattern shown in Figs.\ref{fig:c3-s1s3-msd}(c)
and (f) for $M\geq 3$. 
All the above results allows us to regarded the normal diffusion $m_2(t)\propto t$ as an
ultimate limit of the critical subdiffusion for $M=2$, and $\deff=M=2$ is just the 
critical dimension of localization exhibited by the A-model.

Furthermore, we confirmed that the above features do not change 
when the radom potential $V(n,t)$ is replaced by
\beq
V(n,t)=V_1(n)+V_2(n)f_\eps(t),
\eeq
where $V_1(n)$ and $V_2(n)$ are different random sequences.
The same is true if a binary random sequences taking $-W$ or $W$.
are used for $V_1(n)$ and $V_2(n)$.

\vspace{0.8cm}

\section{Delocalized states}
\label{sec:delocalized states}
In this section, we investigate the characteristics of 
the delocalized states which emerges for $M\geq 3$ and $\eps>\eps_c$ 
in comparison with the stochastic model.
Results are compared also with the B-model with no static random potential.

\subsection{Comparison with stochastic models}
We investigate the dependency upon the two parameters $W$ and $\eps$ in comparison
with the $D$ of the stochastic model by Haken and others 
\cite{haken72,palenberg00,moix13,knap17}.

\begin{figure}[htbp]
\begin{center}
\includegraphics[width=7.0cm]{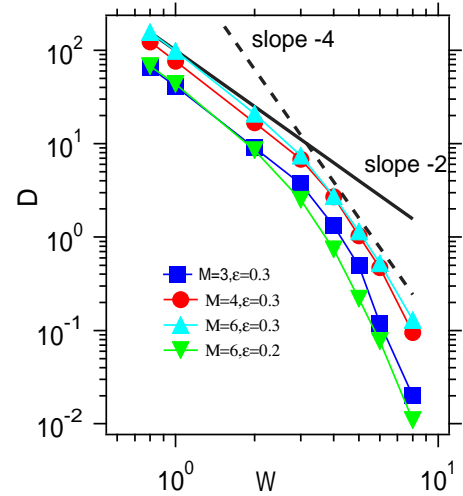}
\caption{\label{fig:dif-con-W}(Color online)
The diffusion coefficient $D$ of the quantum diffusion 
 as a function of $W$ 
in the A-model with $\eps=0.2$ or $\eps=0.3$ of  $M=3,4,6$.
Note that the axes are in the logarithmic scale. 
$D \propto W^{-2}$ and $D \propto W^{-4}$ are shown by black line and 
black dotted lines, respectively, for reference.
}
\end{center}
\end{figure}

Typical examples of the $m_2(t)$ for $\eps>>\eps_c$
in the cases of  $M=3$ and $M=7$ are shown in Fig.\ref{fig:dif-con-W}(a) and (b), respectively. 
If $\eps$ is large enough, it is evident that MSD follows asymptotically
the normal diffusion $m_2=Dt$, which implies that only  finite number of 
coherent modes plays the same role as the stochastic perturbation.

\begin{figure}[htbp]
\begin{center}
\includegraphics[width=9.0cm]{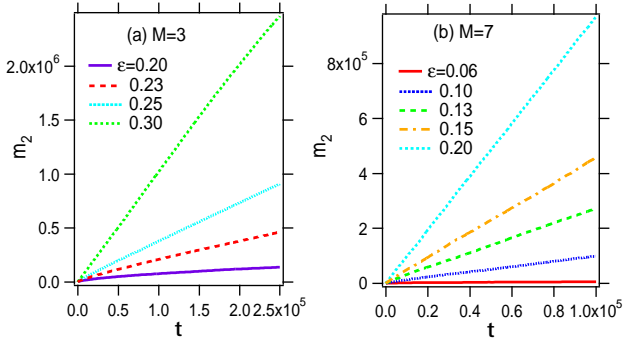}  
\caption{\label{fig:dif-msd-W}(Color online)
The $m_2(t)$ as a function of time in the A-model of  (a)$M=3$ and (b)$M=7$ with $W=1$ 
for some values of the perturbation strength $\eps$,
 increasing from $\eps=0.1$(bottom) to $\eps=0.2$(top) for $M=7$
and from $\eps=0.2$(bottom) to $\eps=0.3$(top) for $M=3$, respectively. 
Note that the axes are in the real scale.
}
\end{center}
\end{figure}

Indeed, the $W-$dependence of the diffusion coefficient $D$ depicted
in Fig.\ref{fig:dif-con-W} follows the main feature of 
the stochastically induced diffusion constants regardless of the number of colors $M(\geq 3)$.
The dependence changes in the weak regime and strong regime of $W$ as 
\beq
 D \propto
\begin{cases}
W^{-2} & (W\ll 1)\\
W^{-4} & (W\gg 1).
\end{cases} 
\eeq
The weak regime result follows Eq.(\ref{eq:D-stochastic}) if $W \gg \eps$. 
The strong regime behavior agrees with  the result obtained 
by Moix {\it et al} \cite{moix13}
for the stochastic model in the very large limit of $W$.

\begin{figure}[htbp]
\begin{center}
\includegraphics[width=7.0cm]{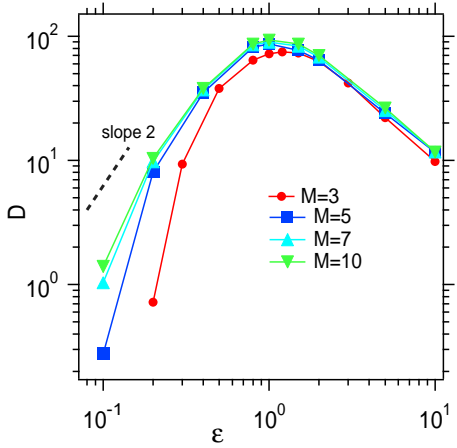}  
\caption{\label{fig:t-cont-st}(Color online)
The diffusion coefficient $D$ 
as a function of  $\eps$ 
for the A-model with $W=1$ and  $M=3,5,7,10$.
Note that the axes are in the logarithmic scale. 
$D \propto \eps^{2}$ is shown by black line for reference.
}
\end{center}
\end{figure}

Next, we examine the $\eps$-dependence of $D$, which is shown in Fig.\ref{fig:t-cont-st}
for some $M$s. 
As a whole, the $\eps$-dependence almost follows Eq.(\ref{eq:D-stochastic}) for all $M$.
(We note that Eq.(\ref{eq:D-stochastic}) is valid for small $\eps$ and $W$, and it can not be 
directly be applied to the interpretation of our result.)
If $\eps$ is weak $D$ increases as 
\beq
\label{eq:Deps1}
  D \propto \eps^2
\eeq
for $M\gg 1$ in agreement with Eq.(\ref{eq:D-stochastic}), 
and after going over the maximum value at $\eps^*\sim O(1)$,
it decreases. In particular in the regime $\eps>\eps^*$, $D$ has no significant $M$-dependence.
This fact implies a remarkable feature that the diffusion induced by the coherent perturbation 
composed of only three incommensurate frequencies mimics the normal diffusion induced 
by a stochastic perturbation containing infinite number of colors.

\subsection{Comparison with the B-model}
\begin{figure}[htbp]
\begin{center}
\includegraphics[width=4.5cm]{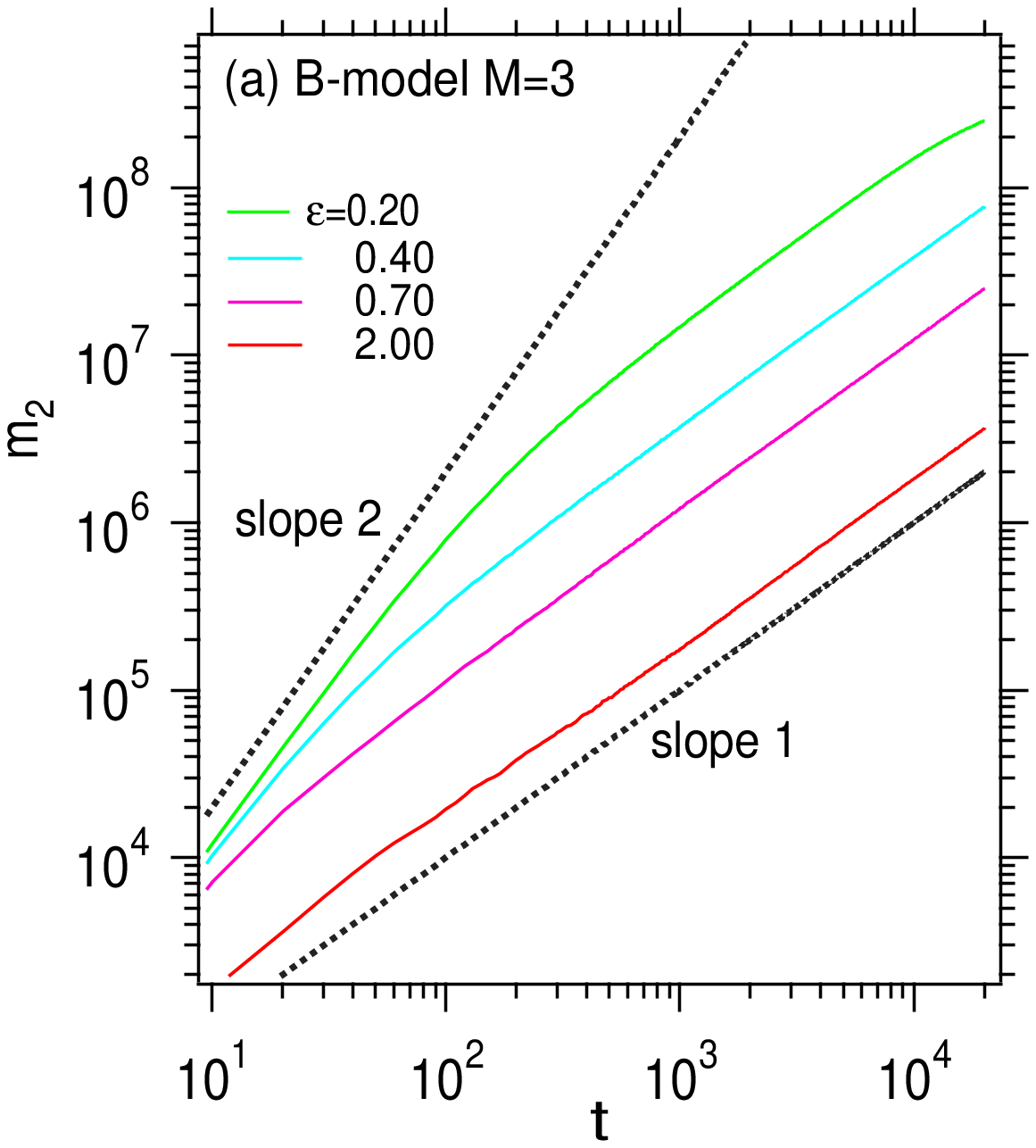}
\hspace{-6mm}
\includegraphics[width=4.5cm]{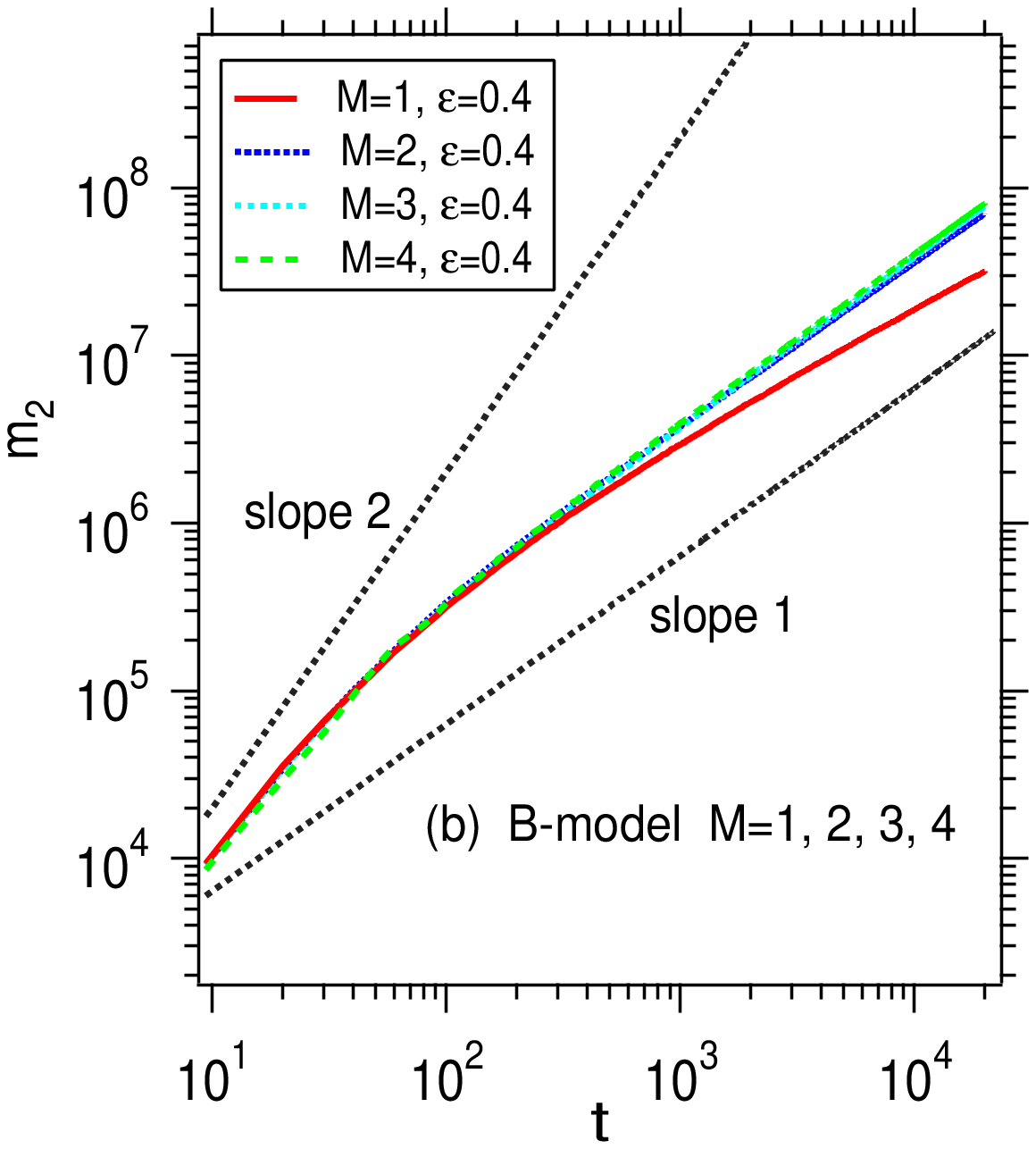}
\caption{\label{fig:dif-msd-W-B-model} (Color online)
The double-logarithmic plots of 
$m_2(t)$ as a function of $t$  in the B-model  with  $W=1$. 
(a)$M=3,W=1$. (b)$M=1,2,3$ and $W=1$.
Note that the axes are in the logarithmic scale. 
Black dotted line shows $m_2(t) \propto t^{1}$ for reference.
}
\end{center}
\end{figure}

In the case of $\eps=0$, the B-model  becomes 
spatially periodic system without potential part, and the wavepacket exactly shows
ballistic motion as $m_2(t) \propto t^2$.
We consider the MSD for finite $\eps$ in the B-model in comparison with the A-model.
Figure.\ref{fig:dif-msd-W-B-model}(a) shows the time evotution of the MSD 
of the B-model with $M=3$ for some values of $\eps$. We can see the ballistic growth 
$m_2(t)\sim t^2$ in the short time regime in the all cases.
As seen in  the $M-$dependence in the Fig.\ref{fig:dif-msd-W-B-model}(b), 
in the B-model of $M=1$, the wavepacket localizes. In contrast, for $M \geq 2$ the normal 
diffusive behavior $m_2\propto t$,
which loses significant $M$-dependence, 
appears as time proceeds. 
For more detailed features of MSD of the B model, see
Appendix \ref{app:B-model-MSD}.

\begin{figure}[htbp]
\begin{center}
\includegraphics[width=7.0cm]{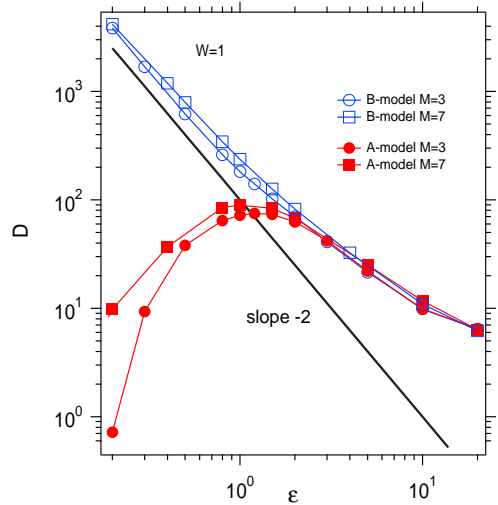}
\caption{\label{fig:t-cont-1}(Color online)
The diffusion coefficient $D$ of the quantum diffusion 
 as a function of  $\eps$ 
in the A-model and B-model for several $M$ with $W=1$.
The corresponding results for the ballistic model are also provided.
Note that the axes are in the logarithmic scale. 
$D \propto \eps^{-2}$ is shown by black line for reference.
}
\end{center}
\end{figure}

Figure \ref{fig:t-cont-1} compares the $\eps-$dependence of 
the diffusion coefficients $D$ of the B-model with those of the A-model.
The difference between A-model and B-model is evident in the region $\eps_c<\eps<\eps^*$.
In A-model, as was stated above, $D$ increases first like $\eps^{2}$ in $\eps<\eps^*$
and it decreases beyond $\eps^*$. But in B-model $D$ decreases monotonously. 
In the regime $\eps<\eps^*$, $D$ decreases in contrast to Eq.(\ref{eq:Deps1}) as
\beq
       D \propto \eps^{-2}
\eeq
Beyond  $\eps^*$, $D$ continues to decreases, which is closely followed  by A-model.
Thus the diffusion processes of the two models become indistinguishable in the region $\eps\gg\eps^*$ for $M\geq 3$. 
 The above tendency is the same even when we examine the
the stochastic model by replacing $f_{\eps}(t)$ with $n(t)$.
For $\eps > \eps^*$, the $\eps-$dependence of the diffusion coefficient
$D$ of the SA-model also approach those of the SB-model.
(See Fig.\ref{fig:dif-con-W-SASB}
 in Appendix \ref{app:B-model-MSD}.)

\section{Summary and discussion}
\label{sec:summary}
We investigated systematically the localization-delocalization 
transition (LDT) of the one-dimensional Anderson model which 
is dynamically perturbed by polychromatically
quasi-periodic oscillations by changing the three parameters;  
the disorder strength $W$, perturbation strength $\epsilon$ and 
the number of the colors $M$ of the oscillations.
The dynamical localization length (LL) was evaluated
by the MSD computed by the numerical wavepacket propagation. 
Although our model consists of $M+1$ degrees of freedom,
we analyzed the numerical results under the hypothesis
that the effective dimension $d_f$ is $M$, not $M+1$, 
considering the energy conservation.
The transition to delocalization is observed for $M=d_f+1 \geq 3$,
and for $M=d_f+1\leq 2$ only localization takes place, 
which are consistent with the $d-$dimensional Anderson model
if $d_f$ is identified with $d$.

For $M \leq 2$ the LL increases exponentially with respect
to $\epsilon$ if $\eps$ is relatively small. 
On the other hand, 
the $W-$dependence of the LL is also scaled by the disorder strength $W$ as in
the case of Anderson map (AM).

For $M \geq 3$ the localization-delocalization transition(LDT) always takes place 
with increase in the perturbation strength $\eps$, and  at the critical point $\eps_c$ 
the fractional diffusion MSD$\propto t^\alpha$~($0<\alpha<1$) is observed.
The critical diffusion exponent decreases as $\alpha \simeq 2/M$ with $M$ in accordance with 
the prediction of one-parameter scaling theory (OPST) 
under the hypothesis $d_f=M$.
The numerical results reveal that the critical perturbation strength decreases as 
$\eps_c\propto 1/(M-2)^{1/2}$  with an increase of $M$.
These properties are different from those of the AM system  
reported in the previous papers \cite{yamada20}. 
On the other hand, 
the dimensional dependence of the critical exponent $\nu$ of 
the localization length (LL) roughly estimated by the numerical data  
was qualitatively consistent result with those of 
the polychromatically perturbed AM system with $(M-1)$ colors and 
the LDT in  $d-$dimensional Anderson model.

The Table \ref{fig:table1} summarizes the localization and delocalization 
phenomena of the random systems, 
including the case of the random system of the spatial dimension $d$ 
and the perturbed quantum map systems.

We also studied the delocalized states for $\eps>\eps_c$.
Even though $M$ is not large,
the $W-$ and $\eps-$dependence of the diffusion coefficient of 
the delocalized states mimics those predicted for the stochastically
perturbed 1D Anderson model.

As $\eps > O(1)$ the characteristics of diffusion of our model approaches 
closely to those of the B-model which contains only the quasi-periodically
oscillating random potential and has no static randomness.

\begin{table}[htbp]
\begin{center}
 \caption{\label{fig:table1}
 Dimensionality of the DLT.
For $4 \leq M <\infty$ the result is same as the case of $M=3$.
The lower lines is result of the $d-$dimensional disordered systems 
by the scaling theory of the localization.
Loc: exponential localization, LDT:localization-delocalization transition, 
Diff:Normal diffusion.
}
 \begin{tabular}{cccccc}
\hline
\hline
$M$ &0 & 1 & 2 & 3 & 4 \\ \hline 
this study(A-model)  & Loc & Loc & Loc & LDT & LDT \\ 
this study(B-model)  & Bali & Loc & Diff & Diff & Diff \\
Anderson map \cite{yamada20}  & Loc & Loc & LDT & LDT  &LDT \\
Standard map \cite{yamada20}  & Loc  & Loc & LDT & LDT  &LDT\\ 
 \hline
 d &  1 & 2 & 3 &4 &5\\ \hline 
Anderson model    & Loc & Loc & LDT & LDT & LDT\\ \hline
 \end{tabular}
\end{center}

 \end{table} 

\appendix

\section{Frequency set used in the calculation}
\label{app:omega}
Table \ref{table:omega-set} shows the sets, $S_1$,$S_2$,$S_3$, 
 of  the frequency set $\{ \omega_i \}$. 
$S_1$ is mainly used in the text, and 
as mentioned in the text, which is set to be $O(1)$ 
in the incommensurate as much as possible.
The frequency set relatively  
 affects the numerical result compared to the case of the Anderson map system, 
although the larger the $M$, the smaller the influence of how to select the frequency.
Therefore, in addition to the fundamental frequency set $S_1$, we investigated 
the result in the A-model 
with the other frequency sets $S_2$, $S_3$ given in the Table \ref{table:omega-set}.
Randomly chosen values are used for $S_3$.
$S_2$ was used for  numerical calculation by 
6th order symplectic integrator in our previous paper \cite{yamada99}.

\begin{table}[htbp]
\begin{center}
\caption{\label{table:omega-set}
The frequencies $S_1$ we mainly used are followings: 
$\omega_1=(1+\sqrt{5})/2$, $\omega_2=2\pi/\lambda$, $\omega_3=2\pi/\lambda^2$,
$\omega_4=\sqrt{3}-1$, $\omega_5=\sqrt{2}-1$, $\omega_6=\sqrt{13}/2-1$,
$\omega_7=\sqrt{11}-3$, $\omega_8=\sqrt{10}/2-1$, $\omega_9=5\sqrt{17}-20$,
$\omega_{10}=2\sqrt{19}/2-1$, 
where $\lambda$ denotes the real root of the cubic equation $x^3-x-1=0$.
We have checked for another set of the frequencies.
The whole tendency of the main result in the present paper is not depend on
the choice for the long-time calculation with large system size.
$S_2$ and, $S_3$ are used to get the data of $M=6$ and $M=7$ for check.
$r_k$($k=1...,7$) take  uniform random number within $[0,1]$.
}
    \begin{tabular}{llcc}
  \hline \hline
$\omega_M$ & $S_1$   & $S_2$ & $S_3$ \\ \hline
$\omega_1$   & $\sigma$  & 1+$\sqrt{1/7}$  & 1/2+$r_1$  \\
$\omega_2$   & $\nu_1$  & 1+$\sqrt{2/7}$  & 1/2+$r_2$        \\
$\omega_3$   & $\nu_2$ & 1+$\sqrt{3/7}$  & 1/2+$r_3$       \\
$\omega_4$   & $\sqrt{3}-1$   &  1+$\sqrt{5/7}$  & 1/2+$r_4$        \\
$\omega_5$   & $\sqrt{2}-1$ &  1+$\sqrt{7/7}$ & 1/2+$r_5$        \\
$\omega_6$   & $\sqrt{13}/2-1$  &  1+$\sqrt{10/7}$ & 1/2+$r_6$        \\
$\omega_7$   & $\sqrt{11}-3$ &  1+$\sqrt{11/7}$  & 1/2+$r_7$        \\
$\omega_8$   & $\sqrt{10}/2-1$ &  ---  & ---        \\
$\omega_9$   & $5\sqrt{17}-20$ &  ---  & ---       \\
$\omega_{10}$   & $2\sqrt{19}/2-1$ &  ---  & ---       \\
\hline
  \end{tabular}
\end{center}
\end{table}
%
%

\section{Autonomous representation of the time-dependent 
Schr\"{o}dinger equation (\ref{eq:model})}
\label{app:auto}
Let the wavefunction describing the whole system composed of 
the one-dimensional lattice and the $M$ harmonic modes be $|\Psi(t)\>$.
We introduce the set of the action-angle operators 
$(\hat{J}_i,\hat{\phi}_i):=(-i\hbar \frac{\pr}{\pr \phi_i},\phi_i)~~(i=1,2..M)$
representing the harmonic modes,
and let $\hat{H}_0$ be the part of Hamiltonian in Eq.(\ref{eq:model}) 
without the harmonic perturbations
(i.e. $\eps=0$) and 
introduce the Hamiltonian $\hat{h}=\sum_{i=1}^M \omega_i\hat{J}_i$ 
representing the harmonic modes.
The autonomous version of Eq.(\ref{eq:model}) is written as
the evolution equation:
\beq
\label{eq:automodel0}
i\hbar \frac{|\partial \Psi(t)\>}{\partial t}=\hat{H}_{\rm tot}|\Psi(t)\>
\eeq
of the whole system with the total Hamiltonian 
\beq
\label{eq:automodel}
 \hat{H}_{\rm tot} =  \hat{H}_0+\hat{h}+
        \frac{W\eps}{\sqrt{M}} \sum_{N}v_n|n\>\<n|\sum_{i=1}^M\cos(\phi_i), 
\eeq
where 
$H_0=\sum_{n}(|n+1\>\<n|+|n\>\<n+1|)+Wv_n|n\>\<n|=\sum_N E_N|N\>\<N|$
 is the unperturbed Hamiltonian, 
and $|n\>$ is the base specifying the site $n$ of 1D Anderson model. 
The eigenstate of the action operator, which is angle-represented as 
$\<\phi_i|J_i\>=e^{iJ_i\phi_i/\hbar}/\sqrt{2\pi}$ with
the action eigenvalue $J_i=n_i\hbar$, is written as $|n_i\>$, and 
let the eigenstate of isolated one-dimensional lattice $H_0$ be $|N\>$ 
with eigenvalue $E_N$: $H_0|N\>=E_N|N\>$.
By decomposing the quantum state of the total system 
as $|\Psi(t)\>=\sum_{N,\{n_i\}}\Psi(N,\{n_i\})|N,\{n_i\}\>$
Eq.(\ref{eq:automodel}) is rewritten by Eq.(\ref{eq:model}).

Let $\hat{U}_{\rm tot}=\exp\{-i\hat{H}_{\rm tot}t/\hbar\}$ be 
the unitary evolution operator of the total system, and introduce 
the new operator $\hat{U}$ by $\hat{U}_{\rm tot}=\e^{-i\hat{h}t/\hbar}\hat{U}$.
Then the evolution equation
\beq
\nn i\hbar\frac{\partial \hat{U}}{\partial t}
=\left[H_0+\frac{W\eps}{\sqrt{M}} \sum_{n}v_n|n\>\<n|\sum_{i=1}^M\cos(\omega_it+\phi_i)\right]\hat{U}
\label{eq:auto3}
\eeq
is immediately obtained, which is equivalent to Eq.(\ref{eq:model}) if the phase-eigenstate 
$|\phi_1\>, |\phi_2\>,....|\phi_M\>$ is supposed at $t=0$.
The identity
\beq
\label{eq:automodel3}
  \e^{i\hat{J}\omega t/\hbar}\e^{-iK\cos \phi/\hbar}\e^{-i\hat{J}\omega t/\hbar}
     = \e^{-iK\cos(\phi+\omega t)/\hbar}
\eeq
is used.

\section{An alternative representaton of Eq.(\ref{eq:modeld})}
\label{app:M=1dash}
Eq.(\ref{eq:automodel}) allows us to introduce an alternative representation
of Eq.(\ref{eq:modeld}) based upon the quantum state of a single lattice site dressed with
harmonic modes interacting with it.
We demonstrate the $M=1$ case. Let us focus on the part 
of Hamiltonian (\ref{eq:automodel}), 
from which the transfer term $\sum_{n}(|n+1\>\<n|+|n\>\<n+1|)$ is neglected,
\beq
   \hat{H}^{(n)}=\omega \hat{J}+Wv_n(1+\eps\cos\phi)|n\>\<n|,
\eeq
which represents the $n$-site interacting with the harmonic mode.
We set $\omega_1=\omega$.
Suppose its eigenstates of the form $|n,K\>=|n\>|K\>_n$, 
satisfying $\hat{H}_n|n,K\>=E_{n,K}|n,K\>$, where $K$ is the new quantum number 
associates with the
harmonic mode to be introduced later. One can readily find that the $\phi$-representation
$\<\phi|K\>_n:=u_{K,n}(\phi)$ of $|K\>_n$ satisfies the simple equation 
\beq
i\omega \frac{\partial u_{K,n}}{\partial \phi} 
=\left[Wv_n(1+\eps\cos\phi)-E_{n,K}\right]u_{K,n}, 
\eeq
which leads to
\beq
  u_{K,n}(\phi)=\frac{1}{\sqrt{2\pi}}\exp\left[i\frac{(E_{n,K}-Wv_n)\phi 
  - Wv_n\eps\sin\phi}{\hbar\omega}\right],
\eeq
where the quantization condition 
$E_{n,K}-Wv_n=K\omega\hbar$~~($K$ is an arbitrary integer) is required for 
the $2\pi$-periodicity of $u_{K,n}(\phi)$. 
Using the new basis $|n,K\>=|n\>|K\>_n$ we expand the
wavefunction as $\Psi(t)=\sum_{n,K}\Phi(n,K)|n,K\>$, 
and the Sch\"{o}dinger equation in  Eqs.(\ref{eq:automodel0}) and (\ref{eq:automodel})
is rewritten into the following form,  instead of the Eq.(\ref{eq:modeld}) with $M=1$, 
\beq
    &&  i\hbar\frac{d\Phi(n,K)}{dt}= (K\omega\hbar+Wv_n)\Phi(n,K)  \nn \\
 && +\sum_{K'}\left[T_{K-K'}^{n,n+1}\Phi(n+1,K')+T_{K'-K}^{n,n-1}\Phi(n-1.K')\right]. 
 \nn \\
\eeq
Then the effective position dependent hopping is given as 
\beq
T_{K-K'}^{n,n'}:= _n\<K|K'\>_{n'}=J_{K-K'} \left(\frac{\eps W(v_n-v_{n'})}{\hbar\omega}\right), 
\eeq
where $J_n(x)$ is the first kind of Bessel function.
We can see that the monochromatic perturbation combined the randomness 
is completely incorporated into 
the hopping terms.
The amplitude $\Phi(n,K)$ at the each lattice site $(n,K)$  is  connected to those at the sites 
$(n\pm1,K-K')$.
It follows that for $\eps W/\hbar \omega <<1$ the hopping coefficients 
 decay along the $K-$direction, and the system becomes 
quasi-1D tight-binding model because $J_n(x) \sim \frac{x^n}{2^nn!}$ as $n \to \infty$.

Similarly, in the case of the B-model of $M=1$, 
 the model can be converted 
into a tight-binding model without the on-site randomness and with hopping randomness.

\section{Result of finite-time critical scaling analysis}
\label{app:critical-scaling-data}
\begin{figure}[htbp]
\begin{center}
\includegraphics[width=7.0cm]{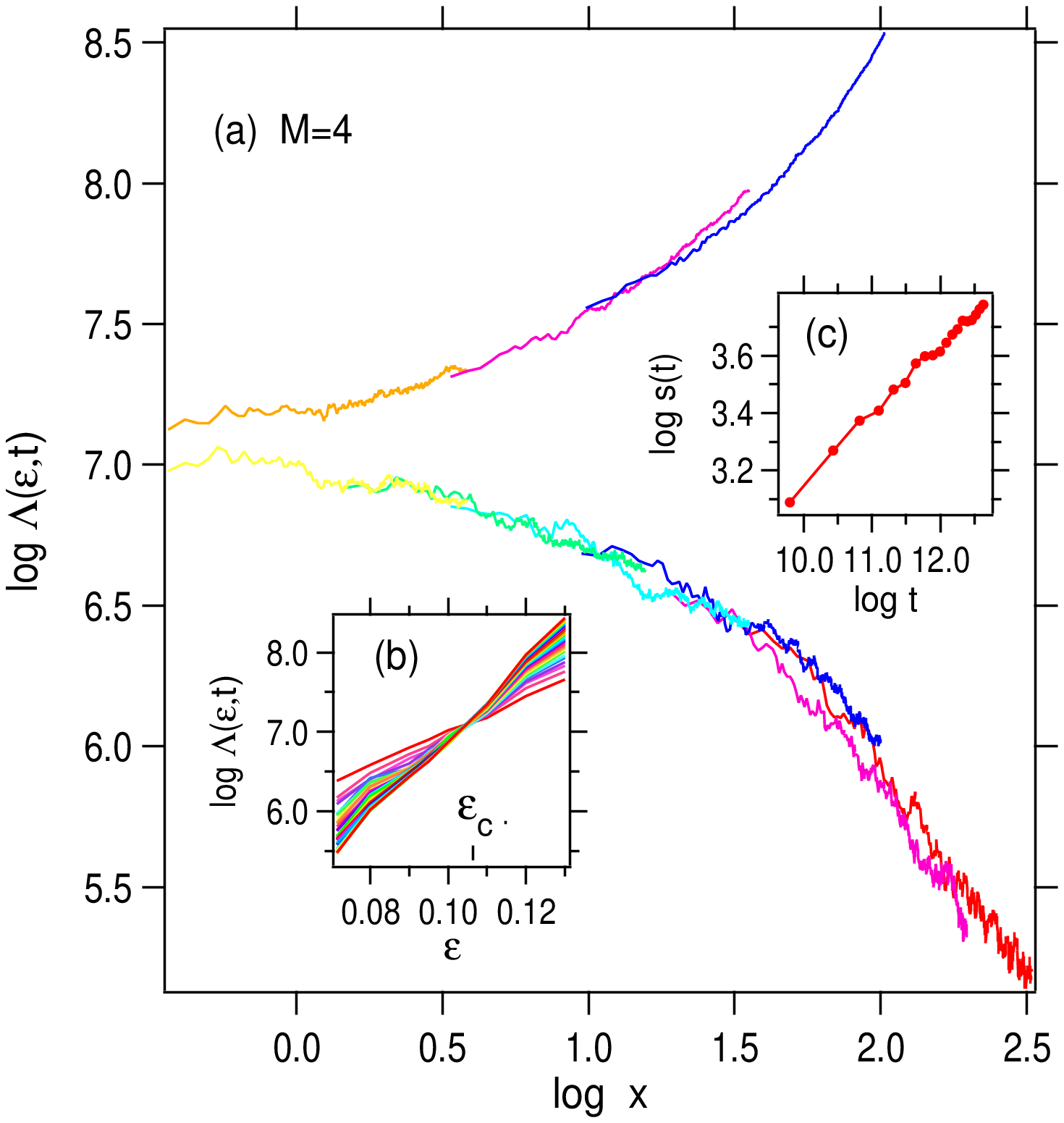}
\caption{\label{fig:c4-nu}(Color online) 
The results of the critical scaling analysis for 
the  A-model  of $M=4$ with $W=1.0$.
(a)The  scaled MSD $\Lambda(\eps,t)$ 
as a function of $x=\xi_Mt^{\alpha/2\nu}$ in the logarithmic scale for some values 
of $\eps$, where $\xi_M$ is the localization length as a scaling parameter .
(b)The scaled $\Lambda(t)$ with $\alpha=0.5$ 
as a function of $\eps$  for some pick up times.
The crossing point is $\eps_c \simeq 0.115$.
(c)$s(t)$ as a function of $t$.
The critical exponent $\nu \simeq 0.9$ is determined by a scaling relation 
Eq.(\ref{eq:s(t)}) by the least-square fit.
}
\end{center}
\end{figure}

\begin{figure}[htbp]
\begin{center}
\includegraphics[width=7.0cm]{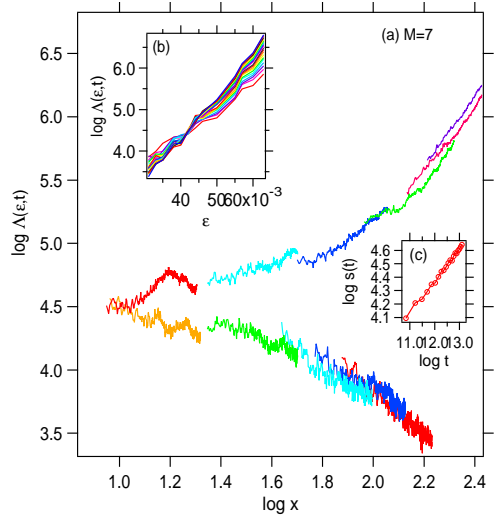}
\caption{\label{fig:c7-nu}(Color online) 
The results of the critical scaling analysis for 
the A-model  of $M=7$ with $W=1.0$.
(a)The scaled MSD $\Lambda(\eps,t)$ 
as a function of $x=\xi_M(\eps)t^{\alpha/2\nu}$  in the logarithmic scale for some values 
of $\eps$, where $\xi_M$ is the localization length as a scaling parameter.
(b)The scaled $\Lambda(t)$ with $\alpha=0.28$ 
as a function of $\eps$  for some pick up times.
The crossing point is $\eps_c \simeq 0.042$.
(c)$s(t)$ as a function of $t$.
The critical exponent $\nu \simeq 0.52$ is determined by a scaling relation 
Eq.(\ref{eq:s(t)}) by the least-square fit.
}
\end{center}
\end{figure}

Figure \ref{fig:c4-nu} and \ref{fig:c7-nu} 
displays the results of the finite-time scaling analysis for 
the A-model of $M=4$ and $M=7$, respectively.
As a result, the OPST is well established for the LDT
regardless of the number of colors $M$ and the disorder strength $W$. 

\section{Normal diffusion  of the B-model}
\label{app:B-model-MSD}
\begin{figure}[htbp]
\begin{center}
\includegraphics[width=4.5cm]{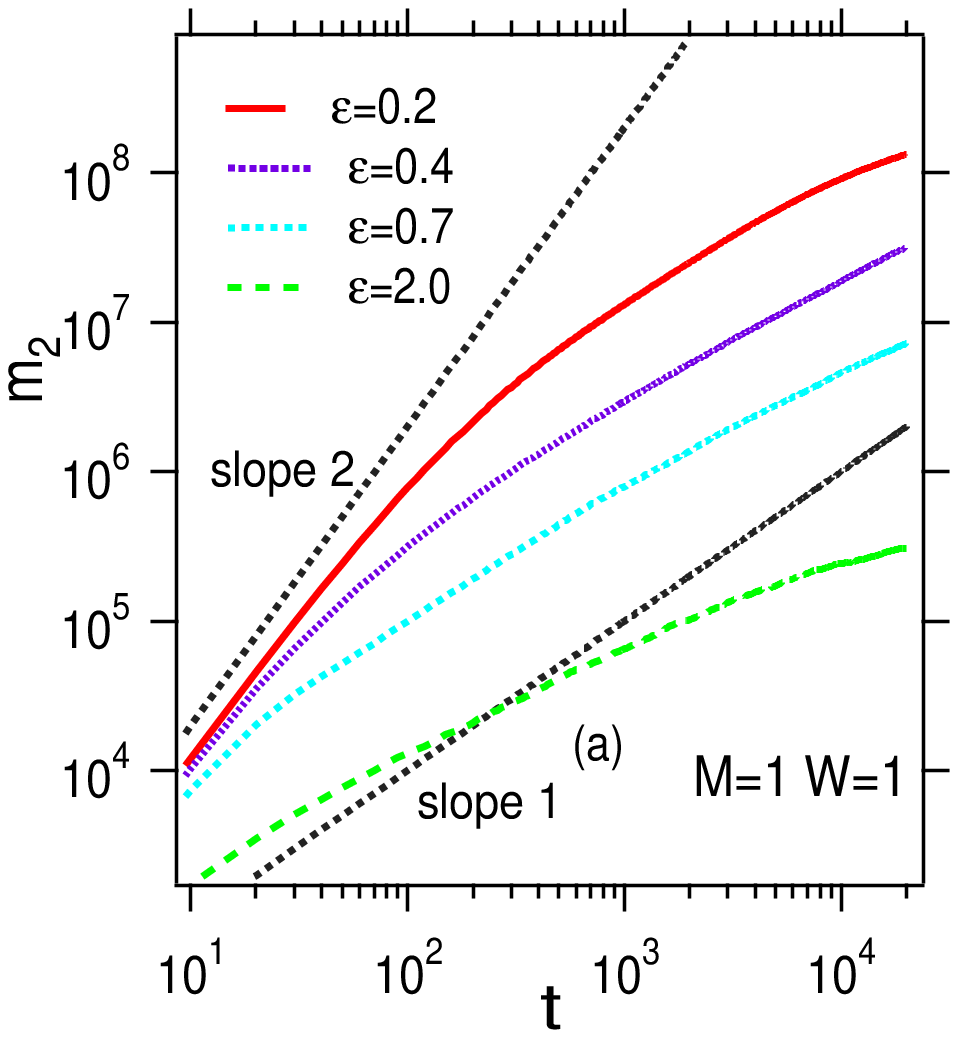}
\hspace{-7mm}
\includegraphics[width=4.5cm]{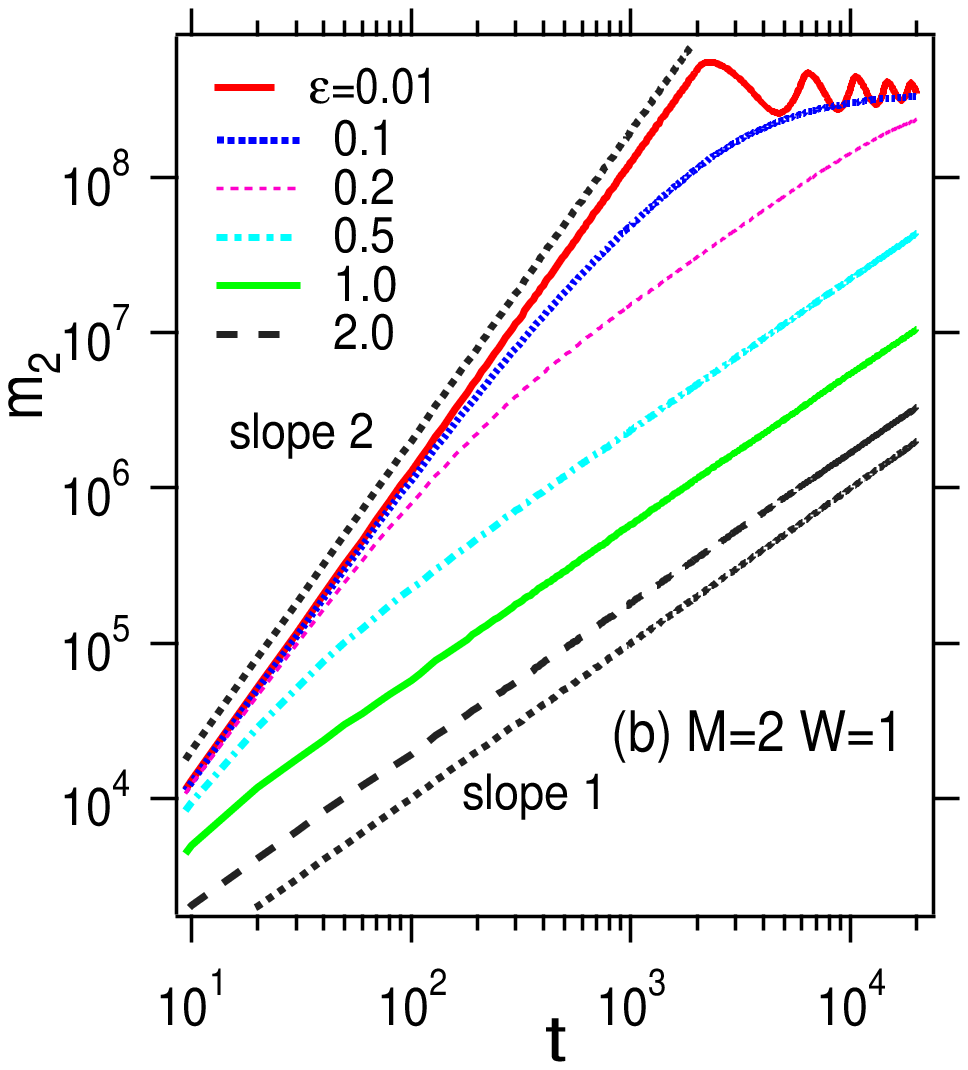}
\caption{\label{fig:ballistic-123}(Color online)
The double-logarithmic plots of $m_2(t)$
 as a function of $t$ for different values of $\eps$ in 
the B-model with  $W=1$. 
 (a)$M=1$ and (b)$M=2$.
Black dotted and thick lines show $m_2(t) \propto t^{2}$ 
and $m_2(t) \propto t^{1}$, respectively, for reference.
}
\end{center}
\end{figure}

Unlike the A-model, in the B-model the system starts with the ballistic
motion $m_2\propto t^2$, and the motion gradually changes as the 
perturbation becomes effective.
The time-dependence of MSD of $M=1$ and $M=2$ is shown in 
 Fig.\ref{fig:ballistic-123}.
As shown in the Fig.\ref{fig:ballistic-123}(a), in the case of $M=1$, 
irrespective of the magnitude of $\eps$, the double-logarithmic 
plots of $m_2(t)$  tells that its instantaneous slope $\alpha_{inst}(t)$ finally 
decreases gradually below $\alpha=1$, and we can not find any 
sign that $\alpha_{inst}(t)$ converges to a non-zero value. 
We, therefore, conjecture that delocalization does not occur for $M=1$.

On the other hand, in the case of $M=2$, as shown in 
Fig.\ref{fig:ballistic-123}(b), the time domain in which the 
ballistic motion is taking place is reduced by increasing $\eps$, and
normal diffusion, $m_2(t) \simeq Dt$, finally appears.
(Due to the system size of numerical calculation, it tends 
to be saturated when it reaches the boundary.)
We conjecture that, no matter how small the magnitude of $\eps$ may
be, the ballistic motion $m_2(t) \simeq t^2$ changes into diffusive motion  
$m_2(t) \simeq t^1$ in a long time limit, if the system size is infinite.
Similar behavior can be expected also for $M\geq 3$, and there is no  LDT.

\begin{figure}[htbp]
\begin{center}
\includegraphics[width=6.2cm]{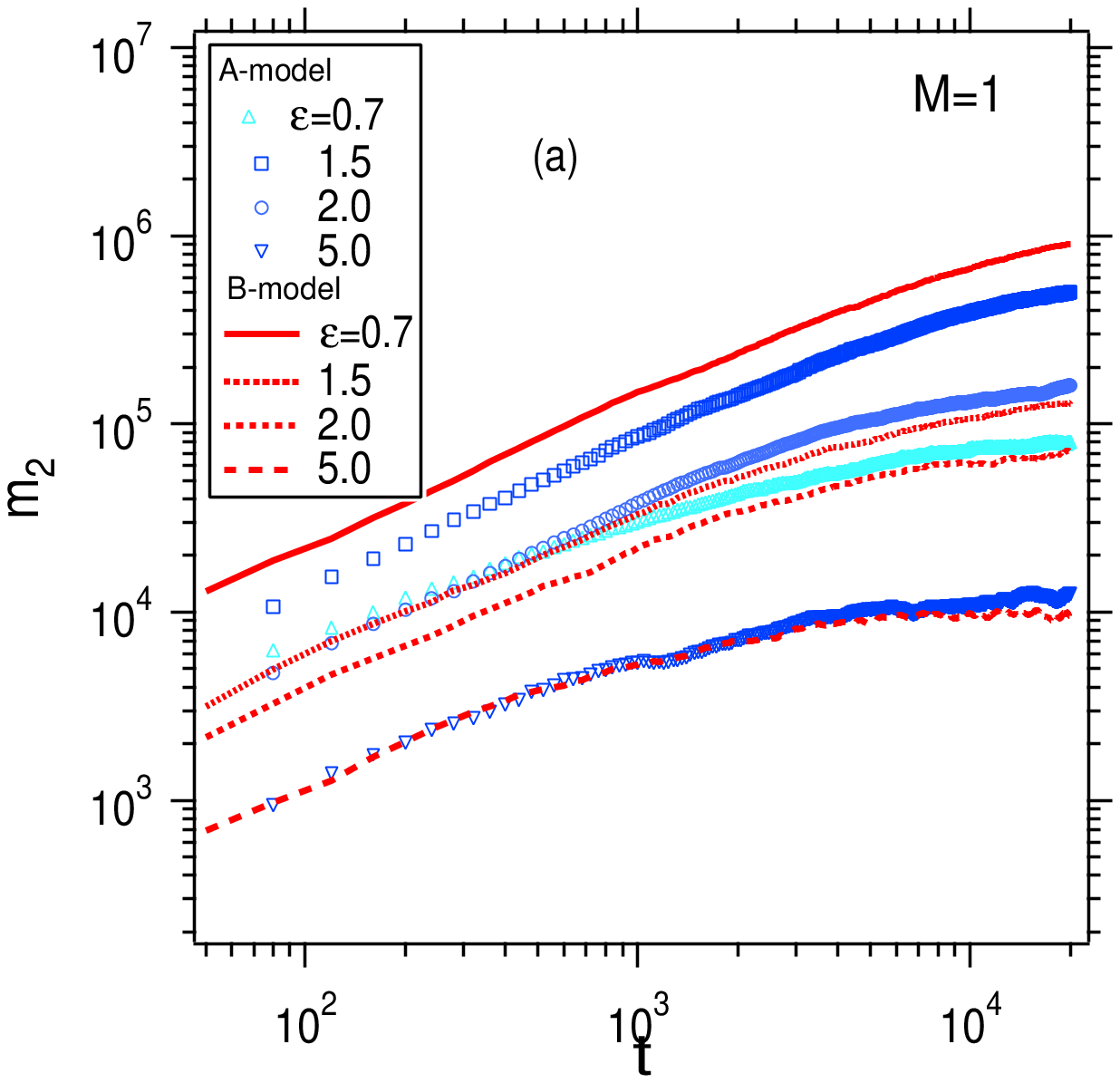}
\hspace{5mm}
\includegraphics[width=6.2cm]{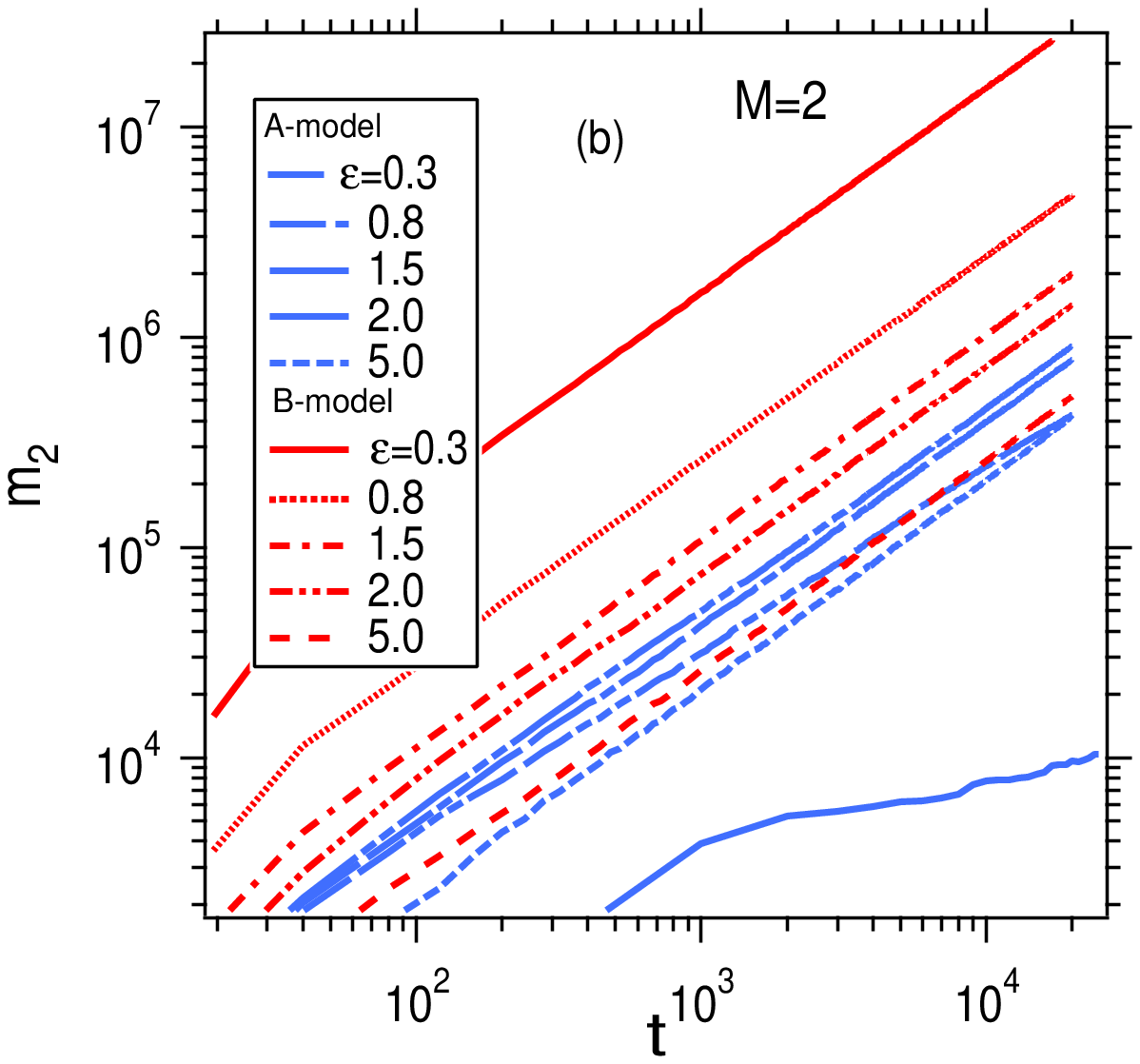}
\caption{\label{fig:t-cont-MSD-2} (Color online)
The double-logarithmic plots of $m_2(t)$
 as a function of $t$ for different values of $\eps$ in 
the A-model and B-model with  $W=1$. 
(a)$M=1$. (b)$M=2$.
}
\end{center}
\end{figure}

Figure \ref{fig:t-cont-MSD-2} shows a comparison of 
 $m_2(t)$ for some $\eps$'s in the A-model and B-model.
Figure \ref{fig:t-cont-MSD-2}(a) is for $M=1$.
The MSD of the A-model increases as $\eps$ increases, 
but it turns to decreases for $\eps>\eps^*$, 
where $\eps^* \sim 1$ is the characteristic value given in the text.
At $\eps>>\eps^*$, it can be seen that the $m_2(t)$ of the A-model 
approaches the result of the B-model, and it overlaps for $\eps=5$
with that of the B-model.
Both cases becomes localized.
As mentioned in the main text, it can be said that it is an asymptotic transition 
from the A-model to the B-model as $\eps$ increases.
Figure \ref{fig:t-cont-MSD-2}(b) is the result for $M=2$.
In the $\eps=5.0$ both cases show normal diffusive behavior for $\eps>>1$.

Moreover, as can be seen in the localized case of the $\eps=0.7$ and $\eps=2.0$ 
in the A-model of $M=1$ in Fig.\ref{fig:t-cont-MSD-2} (a), 
 the time dependence of $m_2(t)$ intersects.
It follows that the two types of regions, $\eps<\eps^*$ and $\eps>\eps^*$, 
do not follow the same scaling curve towards localization 
even if the localization length is the same

\begin{figure}[htbp]
\begin{center}
\includegraphics[width=7.0cm]{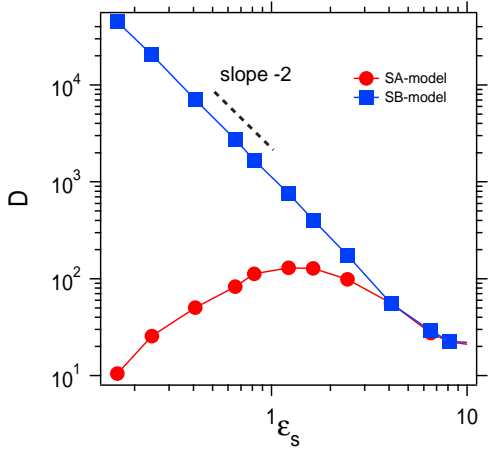}
\caption{\label{fig:dif-con-W-SASB}(Color online)
The diffusion coefficient $D$ of the quantum diffusion 
 as a function of $\eps$ 
in the SA-model and SB-model with $W=2.0$.
Note that the axes are in the logarithmic scale and 
 $\eps_s^* \simeq 1.15$ for $W=2$.
$D \propto \eps_s^{-2}$ is shown by black dotted lines as a reference.
}
\end{center}
\end{figure}

Figure \ref{fig:dif-con-W-SASB} shows a comparison of the  
$\eps_s-$dependence of the diffusion coefficient in the SA-model and SB-model.
It can be seen that the SA-model has a peak around  $\eps_s^*\simeq 1.15$ 
and $D(\eps_s)$ gradually approaches that of the SB-model for $\eps_s>>\eps_s^*$.
This tendency is the same as the relationship between the A-model and the B-model.

\section*{Acknowledgments}
They are also very grateful to Dr. T.Tsuji and  Koike memorial
house for using the facilities during this study.



\end{document}